\documentclass[11pt,a4paper]{article}
\usepackage{amssymb,a4wide}
\usepackage{graphicx}
\usepackage[small]{caption}

\newlength{\dslashwidth}

\newcommand{\tb}{\ensuremath{\tan\beta}}

\newcommand{\bq}{\begin{equation}}
\newcommand{\eq}{\end{equation}}
\newcommand{\ba}{\begin{array}}
\newcommand{\ea}{\end{array}}
\newcommand{\bqa}{\begin{eqnarray}}
\newcommand{\eqa}{\end{eqnarray}}

\newcommand{\lnf}{{\ifmmode \Lambda^{(N_f)} \else $\Lambda^{(N_f)}$\fi}}
\newcommand{\ms}{{\ifmmode \overline{MS} \else $\overline{MS}$\fi}}
\newcommand{\dr}{{\ifmmode \overline{DR} \else $\overline{DR}$\fi}}
\newcommand{\lms}{{\ifmmode \Lambda^{(5)}_{\overline{MS}} \else $\Lambda^{(5)}_{\overline{MS}}$\fi}}
\newcommand{\lam}{{\ifmmode \Lambda \else $\Lambda$\fi}}
\newcommand{\gev}{{\ifmmode {\rm GeV} \else ${\rm GeV}$\fi}}
\newcommand{\gevc}{{\ifmmode {\rm GeV/c^2} \else ${\rm GeV/c^2}$\fi}}
\newcommand{\tev}{{\ifmmode {\rm TeV} \else ${\rm TeV}$\fi}}
\newcommand{\tevc}{{\ifmmode {\rm TeV/c^2} \else ${\rm TeV/c^2}$\fi}}
\newcommand{\lp}{{\ifmmode L^+  \else $L^+$\fi}}
\newcommand{\lm}{{\ifmmode L^-  \else $L^-$\fi}}
\newcommand{\mlp}{{\ifmmode M(L^-) \else $M(L^-)$\fi}}
\newcommand{\mlz}{{\ifmmode M(L^0) \else $M(L^0)$\fi}}
\newcommand{\lz}{{\ifmmode L^0 \else $L^0$\fi}}
\newcommand{\ev}{{\ifmmode GeV/c^2 \else $GeV/c^2$\fi}}
\newcommand{\tri}{{\ifmmode \triangleup \else $\triangleup$\fi}}
\newcommand{\unl}{{\ifmmode U_{lL^0} \else $U_{lL^0}$\fi}}\newcommand{\gL}{{\ifmmode g_L \else $g_{L}$\fi}}
\newcommand{\gR}{{\ifmmode g_R  \else $g_{R}$\fi}}
\newcommand{\gumu}{{\ifmmode \gamma^{\mu} \else $\gamma^{\mu}$\fi}}
\newcommand{\gunu}{{\ifmmode \gamma^{\nu} \else $\gamma^{\nu}$\fi}}
\newcommand{\gdmu}{{\ifmmode \gamma_{\mu} \else $\gamma_{\mu}$\fi}}
\newcommand{\gdnu}{{\ifmmode \gamma_{\nu} \else $\gamma_{\nu}$\fi}}
\newcommand{\stw}{{\ifmmode\sin^2\theta_W \else $\sin^{2}\theta_{W}$ \fi}}
\newcommand{\sws}{{\ifmmode \;\sin^2\theta_W  \else $\;\sin^{2}\theta_{W}$ \fi}}
\newcommand{\cws}{{\ifmmode \;\cos^2\theta_W  \else $\;\cos^{2}\theta_{W}$ \fi}}
\newcommand{\sw}{{\ifmmode \;\sin\theta_W  \else $\sin\theta_{W}$ \fi}}
\newcommand{\cw}{{\ifmmode \;\cos\theta_W  \else $\;\cos\theta_{W}$ \fi}}
\newcommand{\tw}{{\ifmmode \;\tan\theta_W  \else $\;\tan\theta_{W}$ \fi}}
\newcommand{\qq}{{\ifmmode q\overline{q} \else $q\overline{q}$\fi}}
\newcommand{\lR}{{\ifmmode l_R \else $l_R$\fi}}
\newcommand{\lL}{{\ifmmode l_L \else $l_L$\fi}}
\newcommand{\nt}{{\ifmmode \nu_{\tau} \else $\nu_{\tau}$\fi}}
\newcommand{\nuR}{{\ifmmode \nu_R  \else $\nu_R$\fi}}
\newcommand{\nuL}{{\ifmmode \nu_L  \else $\nu_L$\fi}}
\newcommand{\qR}{{\ifmmode g_R  \else $q_R$\fi}}
\newcommand{\qL}{{\ifmmode q_L  \else $q_L$\fi}}
\newcommand{\qRp}{{\ifmmode q_R'  \else $q_{R}$'\fi}}
\newcommand{\qLp}{{\ifmmode q_L'  \else $q_{L}$'\fi}}
\newcommand{\est}{{\ifmmode e^{\bf \ast} \else $e^{\bf \ast}$\fi}}
\newcommand{\lst}{{\ifmmode l^{\bf \ast} \else $l^{\bf \ast}$\fi}}
\newcommand{\must}{{\ifmmode \mu^{\bf \ast} \else $\mu^{\bf \ast}$\fi}}
\newcommand{\taust}{{\ifmmode \tau^{\bf \ast} \else $\tau^{\bf \ast}$ \fi}}
\newcommand{\pperp}{{\ifmmode p_t  \else $p_t$\fi}}
\newcommand{\et}{{\ifmmode E_t  \else $E_t$\fi}}
\newcommand{\xt}{{\ifmmode x_t  \else $x_t$\fi}}
\newcommand{\smumu}{{\ifmmode \sigma_{\mu\mu}  \else $\sigma_{\mu\mu}$ \fi}}
\newcommand{\eg}{{\ifmmode e\gamma  \else $e\gamma$\fi}}
\newcommand{\epem}{{\ifmmode e^+e^-  \else $e^+e^-$\fi}}
\newcommand{\lplm}{{\ifmmode L^+L^-  \else $L^+L^-$\fi}}
\newcommand{\pp}{{\ifmmode p\overline p  \else $p\overline p$\fi}}
\newcommand{\llz}{{\ifmmode L^0\overline{L}^0 \else $L^0\overline{L}^0$\fi}}
\newcommand{\epemt}{{\ifmmode e^+e^- \to  \else $e^+e^- \to$\fi}}
\newcommand{\eb}{{\ifmmode E_{beam}  \else $E_{beam}$\fi}}
\newcommand{\ip}{{\ifmmode pb^{-1}  \else $pb^{-1}$\fi}}
\newcommand{\upm}{{\ifmmode ^{\pm}  \else $^{\pm}$\fi}}
\newcommand{\de}{{\ifmmode ^{\circ}  \else $^{\circ}$ \fi}}
\newcommand{\appr}{{\ifmmode \sim \else $\sim$ \fi}}
\newcommand{\corresp}{{\ifmmode \stackrel{\wedge}{=} \else $\stackrel{\wedge}{=}$ \fi}}
\newcommand{\sqrts}{{\ifmmode \sqrt{s} \else $\sqrt{s}$\fi}}
\newcommand{\zz}{{\ifmmode Z^0  \else $Z^0$\fi}}
\newcommand{\mz}{{\ifmmode M_{Z}  \else $M_{Z}$\fi}}
\newcommand{\mzs}{{\ifmmode M_{Z}^2  \else $M_{Z}^2$\fi}}
\newcommand{\mw}{{\ifmmode M_{W}  \else $M_{W}$\fi}}
\newcommand{\mws}{{\ifmmode M_{W}^2  \else $M_{W}^2$\fi}}
\newcommand{\mh}{{\ifmmode M_{Higgs}  \else $M_{Higgs}$\fi}}
\newcommand{\gt}{{\ifmmode \Gamma_{tot} \else $\Gamma_{tot}$\fi}}
\newcommand{\msusy}{{\ifmmode M_{SUSY}  \else $M_{SUSY}$\fi}}
\newcommand{\msusys}{{\ifmmode M_{SUSY}^2  \else $M_{SUSY}^2$\fi}}
\newcommand{\su}{{\ifmmode SU(3)_C\otimes\- SU(2)_L\otimes\- U(1)_Y \else $SU(3)_C\otimes SU(2)_L\otimes U(1)_Y$\fi}}
\newcommand{\suthree}{{\ifmmode SU(3)_C  \else $SU(3)_C$\fi}}
\newcommand{\sutwo}{{\ifmmode  SU(2)_L\otimes U(1)_Y \else $SU(2)_L\otimes U(1)_Y$\fi}}
\newcommand{\taup} {{\ifmmode \tau_{proton} \else $\tau_{proton}$\fi}}
\newcommand{\as}{{\ifmmode \alpha_{s}  \else $\alpha_{s}$\fi}}
\newcommand{\mgut}{{\ifmmode M_{GUT}  \else $M_{GUT}$\fi}}
\newcommand{\mguts}{{\ifmmode M_{GUT}^2  \else $M_{GUT}^2$\fi}}
\newcommand{\mze} {{\ifmmode m_0        \else $m_0$\fi}}
\newcommand{\mha}{{\ifmmode m_{1/2}    \else $m_{1/2}$\fi}}
\newcommand{\mb} {{\ifmmode m_{b}    \else $m_{b}$\fi}}
\newcommand{\mt} {{\ifmmode m_{t}    \else $m_{t}$\fi}}
\newcommand{\mts} {{\ifmmode m_{t}^2    \else $m_{t}^2$\fi}}

\newcommand{\mtau}{{\ifmmode m_{\tau}  \else $m_{\tau}$\fi}}
\newcommand{\dpp}{{\ifmmode \delta_{pert} \else $\delta_{pert}$\fi}}
\newcommand{\dnp}{{\ifmmode\delta_{non-pert}\else$\delta_{non-pert}$\fi}}
\newcommand{\dew}{{\ifmmode \delta_{\rm EW}\else $\delta_{\rm EW}$\fi}}
\newcommand{\rt}{{\ifmmode R_{\tau}  \else $R_{\tau} $\fi}}
\newcommand{\rz}{{\ifmmode R_{Z}  \else $R_{Z} $\fi}}

\newcommand{\swb}{{\ifmmode \sin^2\theta_{\overline{MS}} \else $\sin^2\theta_{\overline{MS}}$\fi}}
\newcommand{\cwb}{{\ifmmode \cos^2\theta_{\overline{MS}} \else $\cos^2\theta_{\overline{MS}}$\fi}}

\newcommand{\mzero}{\rm m_0}
\newcommand{\mhalf}{\rm m_{1/2}}

\begin{document}

\author{
W. de Boer\footnote{Invited paper at DARK2004, Texas, 3-9 October, 2004}\\
{\it Institut f\"ur Experimentelle Kernphysik, University of Karlsruhe}\\
Postfach 6980, D-76128 Karlsruhe, Germany\\[2mm]
Email: Wim.de.Boer@cern.ch
}

\title{Indirect Evidence for WIMP  Annihilation \\from Diffuse Galactic Gamma Rays}

\maketitle

\begin{abstract}
The EGRET excess in the diffuse galactic  gamma ray data
above 1 GeV shows all the features expected from Dark Matter WIMP
Annihilation: a)it is present and has the same spectrum in all sky
directions, not just in the galactic plane. b) The intensity of the
excess shows the $1/r^2$ profile expected for a flat rotation curve
outside the galactic disc with additionally   an interesting
substructure in the disc in the form of a doughnut shaped ring at 14
kpc from the centre of the galaxy.  At this radius a ring of stars
indicates the probable infall of a dwarf galaxy, which can explain
the increase in DM density.
From the spectral shape of the excess the WIMP mass is estimated to
be between 50 and 100 GeV, while from the intensity the halo profile
is reconstructed. Given the mass and intensity of the WIMP's the mass
of the  ring can be calculated, which is shown to explain the
peculiar change of slope in the rotation curve at about 11 kpc.
These signals of Dark Matter Annihilation are compatible with
Supersymmetry and have a statistical significance of more than
$10\sigma$ in comparison with a fit of the conventional galactic
model to the EGRET data. The statistical significance combined with
all features mentioned above provide an intriguing hint that the
EGRET excess is indeed  a signal from Dark Matter Annihilation.
\end{abstract}
\section{Introduction}
Cold Dark Matter (CDM) makes up 23\% of the energy of the universe,
as deduced from the WMAP measurements of the temperature
anisotropies in the Cosmic microwave Background, in combination with
data on the Hubble expansion and the density fluctuations in the
universe~\cite{wmap}. The Dark Matter (DM) has to be much more
widely distributed than the visible matter, since the rotation
speeds do not fall off like $1/\sqrt{r}$, as expected from the
visible matter in the centre, but stay more or less constant as
function of distance. For  a "flat" rotation curve the DM has to
fall off slowly,  like $1/r^2$, instead of the exponential drop-off
for the visible matter. The fact that the DM is distributed over
large distances implies that its properties must be quite different
from the visible matter, since the latter clumps in the centre owing
to its rapid loss of kinetic energy by the electromagnetic and
strong interactions after infall into the centre.  Since the DM
apparently undergoes little energy loss, it can have at most weak
interactions. In addition its mass is probably large, as deduced
from the formation of stars as soon as a few hundred million years
after the Big Bang. This time scale of star formation could be
deduced from the polarization of the Cosmic Microwave Background,
which is thought to originate from Compton scattering of the CMB  on
the electrons from the ionized plasma in stars\cite{wmap}. Such an
early formation of stars can only be explained, if the DM became
non-relativistic in the early universe and started to cluster by
gravity after decoupling from other particles roughly 10$^{-9}$ s
after the Big Bang. The baryonic matter fell then into these
potential wells of DM after decoupling from the photons 380.000
years after the Big Bang. Given its weak interactions and heavy mass
the DM particles are generically called WIMP's,  Weakly Interacting
Massive Particles.

According to the rules of particle physics  weakly interacting
particles can annihilate, yielding predominantly  quark-antiquark
pairs in the final state, which hadronize into mesons and baryons.
 The stable decay and fragmentation
products are neutrinos, photons, protons, antiprotons, electrons and
positrons. From these, the protons and electrons disappear in the
sea of many matter particles in the universe, but the photons and
antimatter particles may be detectable above the background,
generated by  particle interactions. Such  searches for indirect
Dark Matter detection have been actively pursued, see e.g the review
by Bergstr$\rm \ddot{o}$m\cite{bergstrom} or more recently by
Bertone, Hooper and Silk \cite{Bertone:2004pz}.

 The present analysis on diffuse galactic gamma rays differs from previous ones
by considering simultaneously the complete sky map {\it and} the
energy spectrum, which allows us to constrain both the halo
distribution {\it and} the WIMP mass.
The  WIMP annihilation cross section from cosmology is discussed in
Section 2, while the constraints on the mass and the DM halo profile
from the EGRET excess are discussed in Section 3, followed by  the
expectation from Supersymmetry in Section 4. The
 summary is given in Section 5.

\section{DM annihilation cross Section  from WMAP and photon flux}
In the early universe all particles were produced abundantly and
were in thermal equilibrium through annihilation and production
processes.
%
At temperatures below the mass of the WIMP's the number density
drops exponentially. The annihilation rate $\Gamma=<\sigma v>
n_\chi$ drops exponentially as well, and if it drops below the
expansion rate, the WIMP's cease to annihilate. They fall out of
equilibrium (freeze-out) at a temperature of about $m_\chi/22$
~\cite{kolb} and a relic cosmic abundance remains.

For the case that $<\sigma v>$ is energy independent, which is a
good approximation in case there is no coannihilation, the present
mass density in units of the critical density is given
by~\cite{jungman}: \bq \Omega_\chi h^2=\frac{m_\chi
n_\chi}{\rho_c}\approx (\frac{2\cdot 10^{-27} cm^3 s^{-1}}{<\sigma
v>})\label{wmap}.\eq One observes that the present relic density is
inversely proportional to the annihilation cross section at the time
of freeze out, a result independent of the WIMP   mass (except for
logarithmic corrections). For the present value of $\Omega_\chi
h^2=0.113\pm0.009$ the thermally averaged total cross section at the
freeze-out temperature of $m_\chi/22$ must have been around $2\cdot
10^{-26} {\rm cm^3s^{-1}}$.

From this cross section the differential gamma flux in a direction
forming an angle $\psi$ with the direction of the galactic center
can be calculated:
\begin{equation}
  \phi_\chi(E,\psi)=\frac{\langle \sigma v\rangle}{4\pi} \sum_f \frac{dN_f}{dE} b_f
  \int_{line~ of~ sight}B_l \frac{1}{2}\frac{\langle \rho_\chi^2\rangle}{M_\chi^2} dl_\psi
\label{gammafluxcont}
\end{equation}
where $b_f$ is the branching ratio into the tree-level annihilation
final state, while $dN_f/d E$ is the differential photon yield for
the final state $f$. The WIMP mass density  enters critically in the
prediction for the flux, since the number of WIMP pairs is equal to
$1/2\,\rho_\chi^2/M_\chi^2$.
The factor $B_l$  is the boost factor,
which represents the local enhancement of the number density with
respect to the average by the expected clustering of DM. For the present
analysis $B_l$ is assumed to be the same in all directions $\psi$,
although near the centre of the galaxy the DM clusters may have been tidally disrupted
by the flyby of nearby stars, thus reducing the boost factor towards the centre.
However, this will only modify the density profile near the centre and not affect
the overall analysis.
 Since the
average of $\rho_\chi^2$ can be significantly larger than $\langle
\rho_\chi\rangle^2$  the boost factor can enhance the flux by one or
two orders of magnitude\cite{dokuchaev}.

As mentioned above and discussed further in the
section on Supersymmetry, the dominant final state is always into
quark pairs. These quarks will be mono-energetic, since the
non-relativistic WIMP's annihilate practically at rest. Therefore
one has to consider only one final state and the corresponding gamma
spectrum from mono-energetic quarks is well known from
electron-positron colliders\footnote{The annihilation is
preferentially into heavy b-quarks, which yield a slightly harder
gamma spectrum than the light quarks. This is the spectrum, which
will be used. In case of light quarks the fit to the data would
require a somewhat heavier WIMP mass to obtain the same spectrum.},
so in principle the only free parameters left are the WIMP mass, the
halo profile, i.e. the distribution of the DM density $\rho_\chi$ in
space and the boost factor. The EGRET data are precise enough to
determine these.

\section{Indirect Dark Matter Detection}\label{sec2}

The neutral particles play  a very special role for indirect DM
searches, since they point back to the source.  The charged
particles change their direction by the interstellar magnetic
fields, energy losses and scattering. Therefore the gamma rays
provide a perfect means to reconstruct the intensity (halo) profile
of the DM by observing the intensity of the gamma ray emissions in
the various sky directions. Of course, this assumes that one can
distinguish between the gamma rays from  DMA  the ones from the
background, which is possible because of the different energy
spectra: the gamma rays from the mono-energetic quarks from DMA
produce a significantly harder spectrum than the gammas from nuclear
interaction, which are produced by the interactions between quarks
with a steeply falling power law spectrum ($\propto E^{-2.7}$).
\begin{figure}[t]
\begin{center}
 \includegraphics [width=0.32\textwidth,clip]{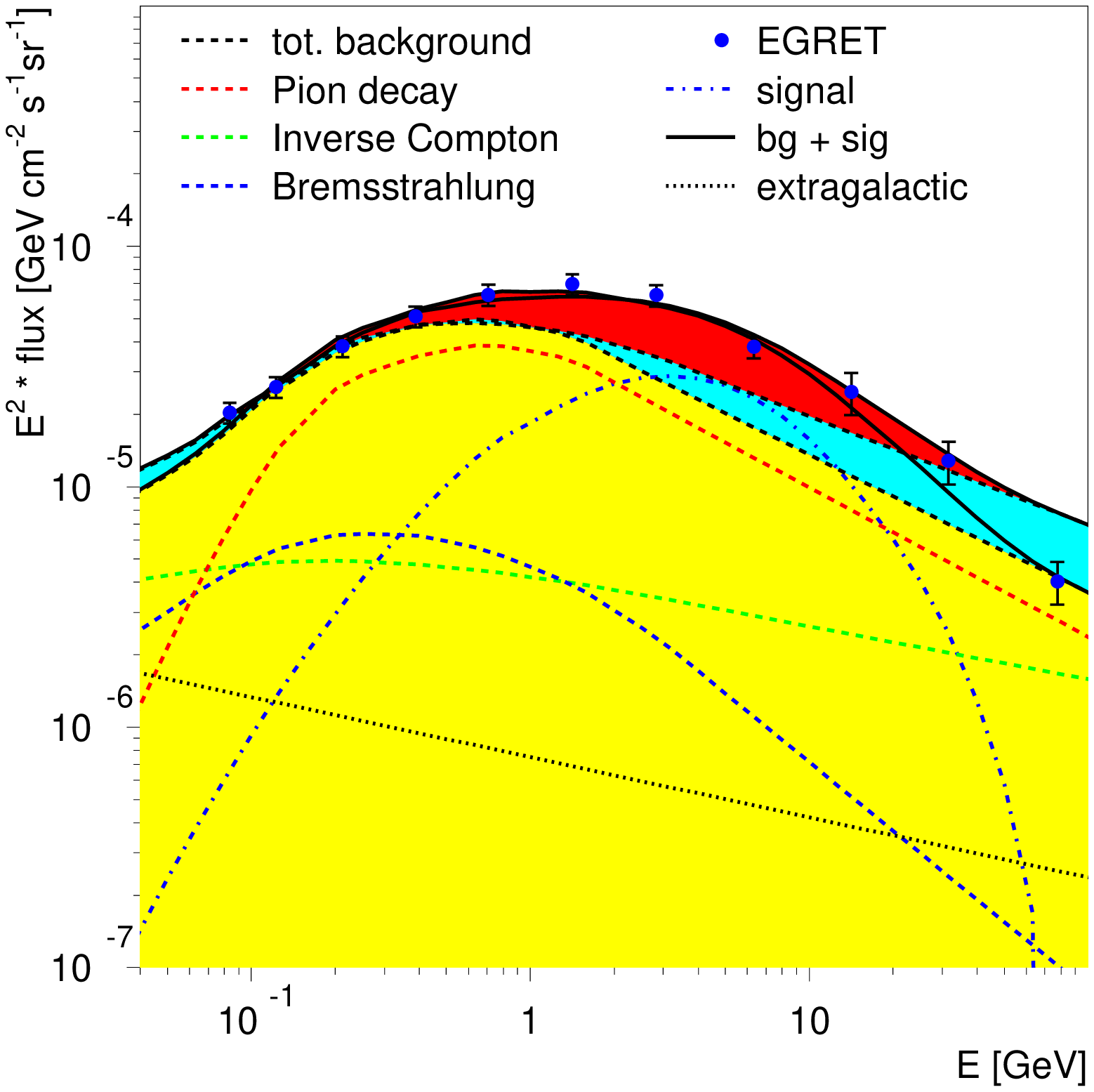}
  \includegraphics [width=0.32\textwidth,clip]{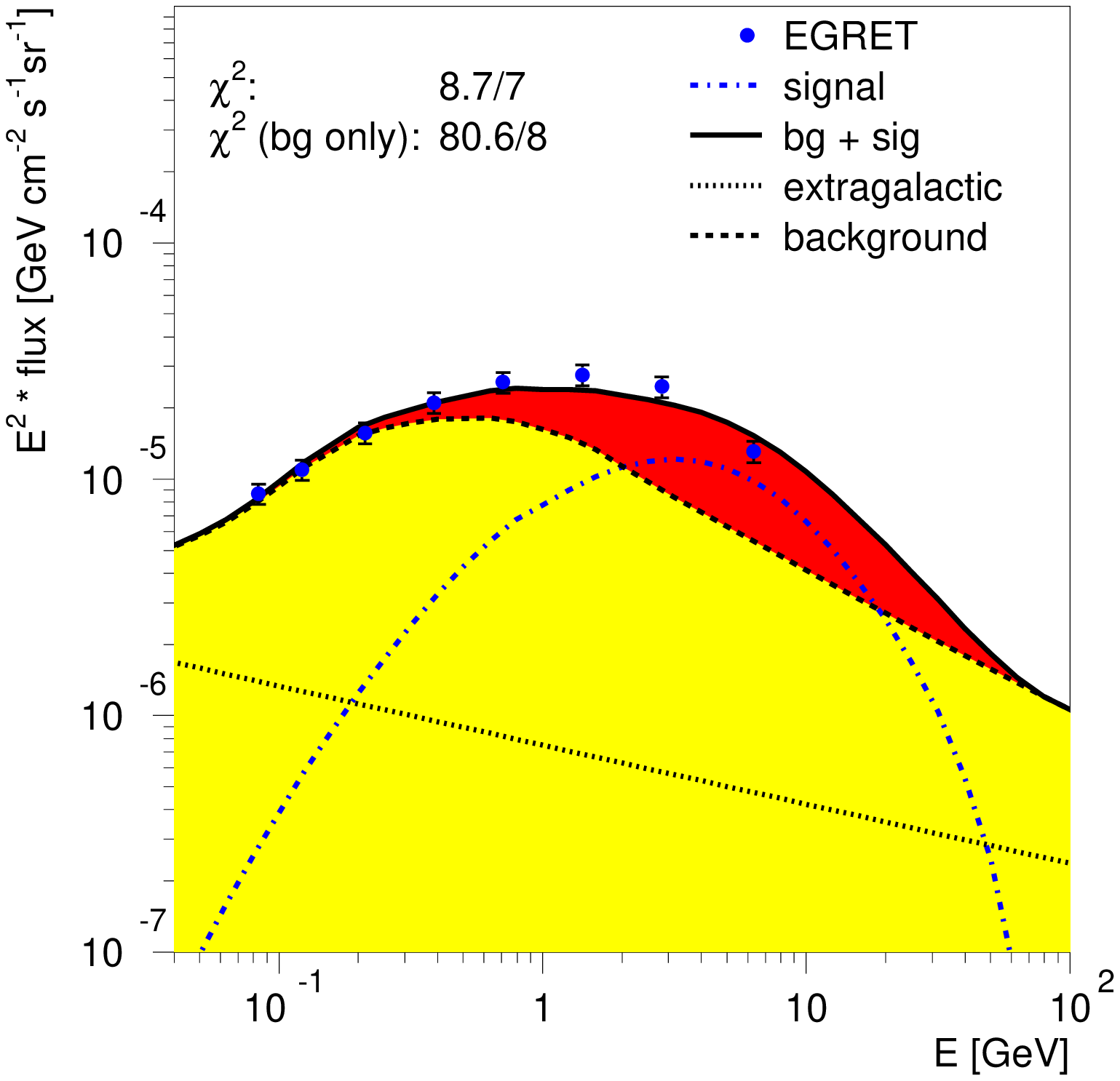}
 \includegraphics [width=0.32\textwidth,clip]{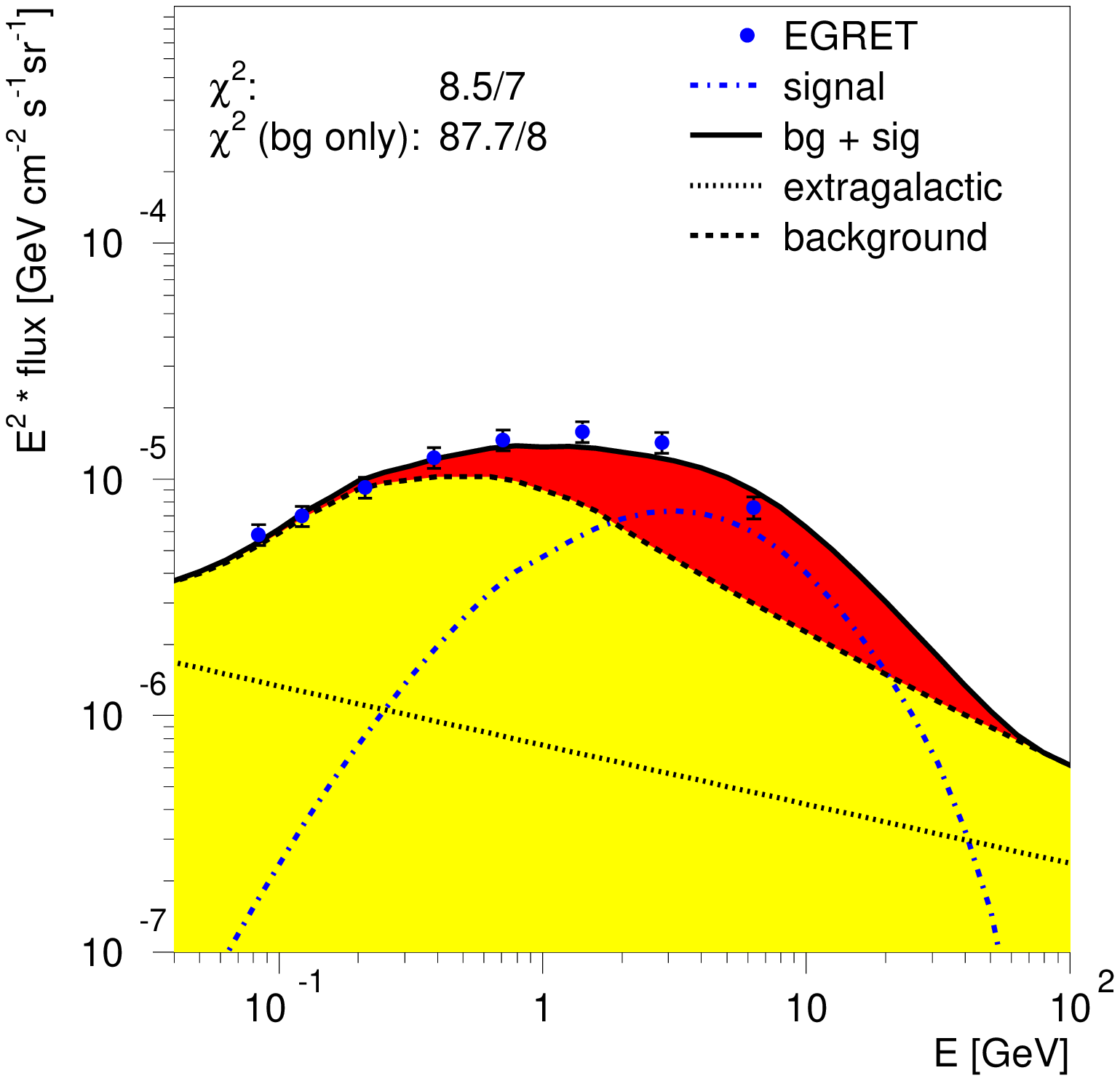}
\includegraphics [width=0.32\textwidth,clip]{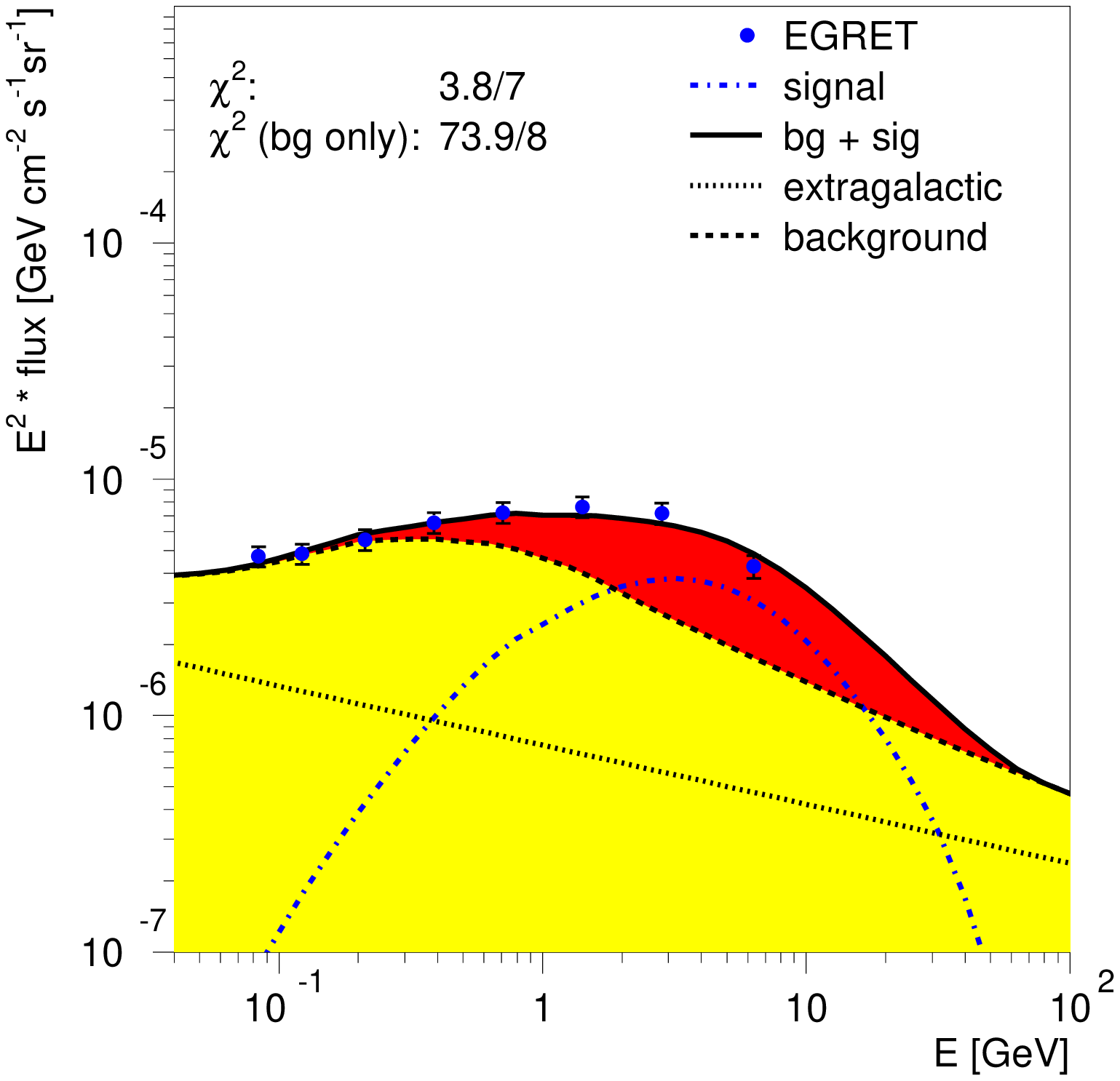}
 \includegraphics [width=0.32\textwidth,clip]{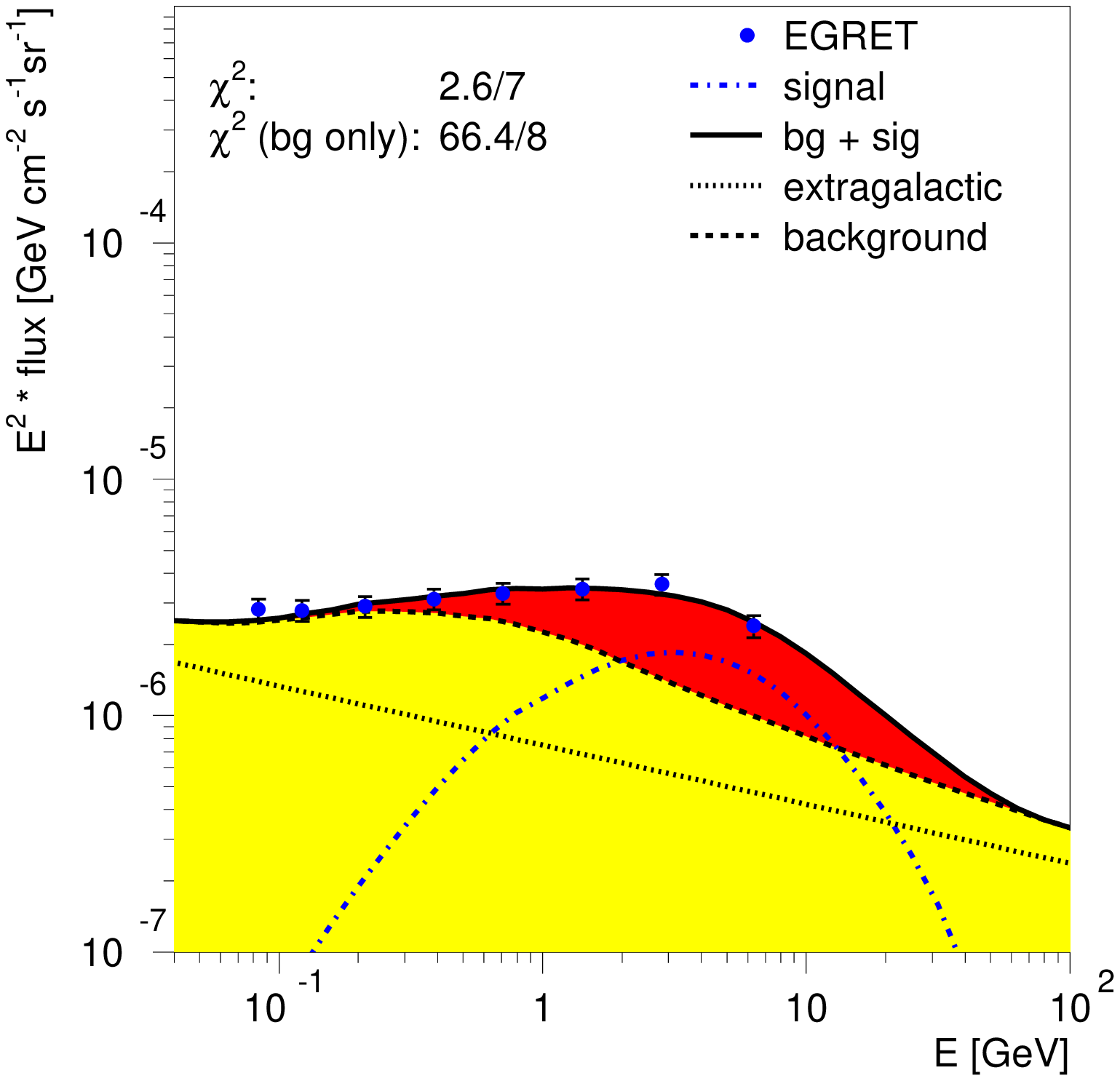}
 \includegraphics [width=0.32\textwidth,clip]{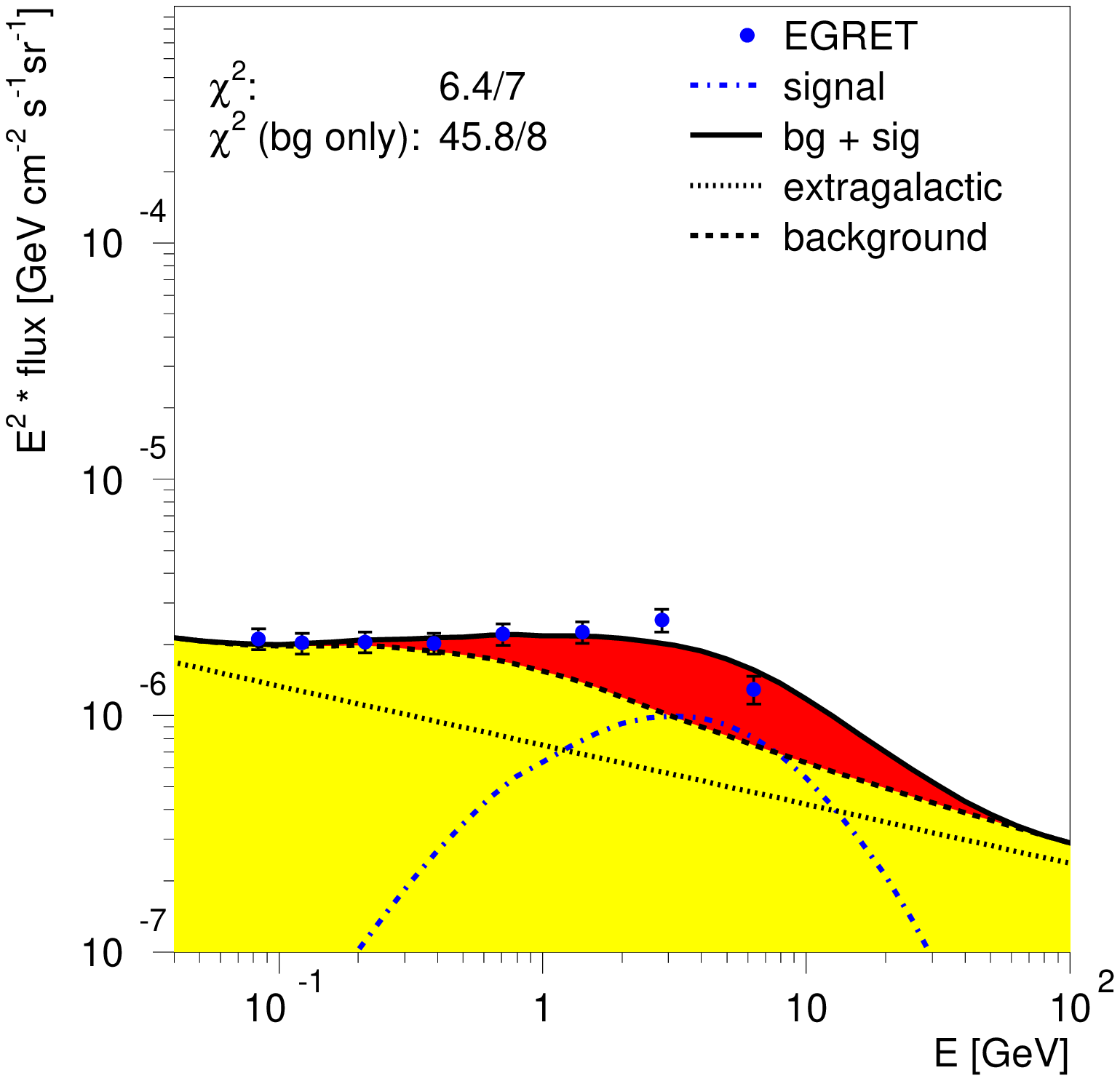}
 \caption[]{
 The diffuse gamma-ray energy spectrum of the disc regions  (top row, from left to right):
  A: towards the galactic centre (latitudes  $0^\circ<|b|<5^\circ$; longitudes $0^\circ<|l|<30^\circ$),
  B: the outer disc (latitudes  $0^\circ<|b|<5^\circ$; longitudes $30^\circ<|l|<330^\circ$),
  C: the galactic anticentre ($0^\circ<|b|<10^\circ$;  $90^\circ<|l|<270^\circ$)
  and outside the disc (bottom, from left to right):
  D: intermediate latitudes I
   ($10^\circ<|b|<20^\circ$;  $0^\circ<|l|<360^\circ$),
  E:  intermediate latitudes II
   ($20^\circ<|b|<60^\circ$;  $0^\circ<|l|<360^\circ$),
  and F: the Galactic pole regions
   ($60^\circ<|b|<90^\circ$;  $0^\circ<|l|<360^\circ$),
  as measured by the EGRET space telescope.
The solid straight line represents
the fitted contribution from the extragalactic background, while the
dotted line indicates the contribution from the annihilation from 65
GeV WIMP's with a boost factor around 70. The  total background
 is indicated by the light (yellow) and the DMA by the  dark (red)  shaded
area, respectively.
   In the top left panel  the various contributions to the
background are indicated as well, while
  the  uncertainties
  from the background  are indicated  by the medium shaded (blue) area.
  One observes that the $\chi^2/d.o.f.$ for the fit including DMA is significantly
  better than the fit for the background only.
 \label{excess}}
\end{center}
\end{figure}

The spectral shape of the gamma rays from either the backgrounds or
the mono-energetic quarks are well known from accelerator
experiments and can be obtained from the well-known PYTHIA code for
quark fragmentation\cite{pythia}.

A very detailed gamma ray distribution over the whole sky was
obtained by the Energetic Gamma Ray Emission Telescope EGRET, one of
the four instruments on the Compton Gamma Ray Observatory CGRO,
which collected data during nine  years, from  1991 to 2000.  The
EGRET telescope was carefully calibrated in the energy range of 0.1
to 30 GeV, but using Monte Carlo  simulations the energy range was
recently extended up to 120 GeV\cite{optimized}  with a
correspondingly larger uncertainty, mainly from the self-vetoing of
the detector by the back-scattering from the electromagnetic
calorimeter into the veto counters for high energetic showers.

It was already noticed in 1997 that the EGRET data showed an excess
in the galactic disk\cite{hunter} of gamma ray fluxes for energies
above 1 GeV if compared with conventional galactic models and
repeated later for all sky directions\cite{optimized}. This analysis
was repeated recently\cite{deboer,deboer1}  using a different
analysis technique on the publicly available EGRET data, namely by
comparing the data not with the absolute fluxes from galactic
models, but only with the shape of the gamma energy spectra from the
galactic background, which is much better known and allows to take
the strongly correlated systematic normalization errors between the
different energy points of the spectrum into account.
 Simultaneously to the galactic background the shapes of    Dark
 Matter Annihilation and the extragalactic background  are fitted.
 Fitting these three contributions yielded
 astonishingly good fits with the free normalization of the background agreeing reasonably
 well with the absolute predictions  of the galactic models\cite{galprop,galprop1}
 for the energies between 0.1 and 0.5 GeV. Above these energies a clear contribution from Dark
Matter annihilation is needed,  but the excess in different sky
directions can be explained by a single WIMP mass and a single boost
factor, as shown in Fig. \ref{excess} for 6 different sky
directions.

Alternative explanations for the excess have been plentiful. Among
them: locally soft electron and proton spectra, implying that in
other regions of the galaxy the spectra are harder, thus producing
harder photon spectra.

\begin{figure}[t]
\begin{center}
 \includegraphics [width=0.45\textwidth,clip]{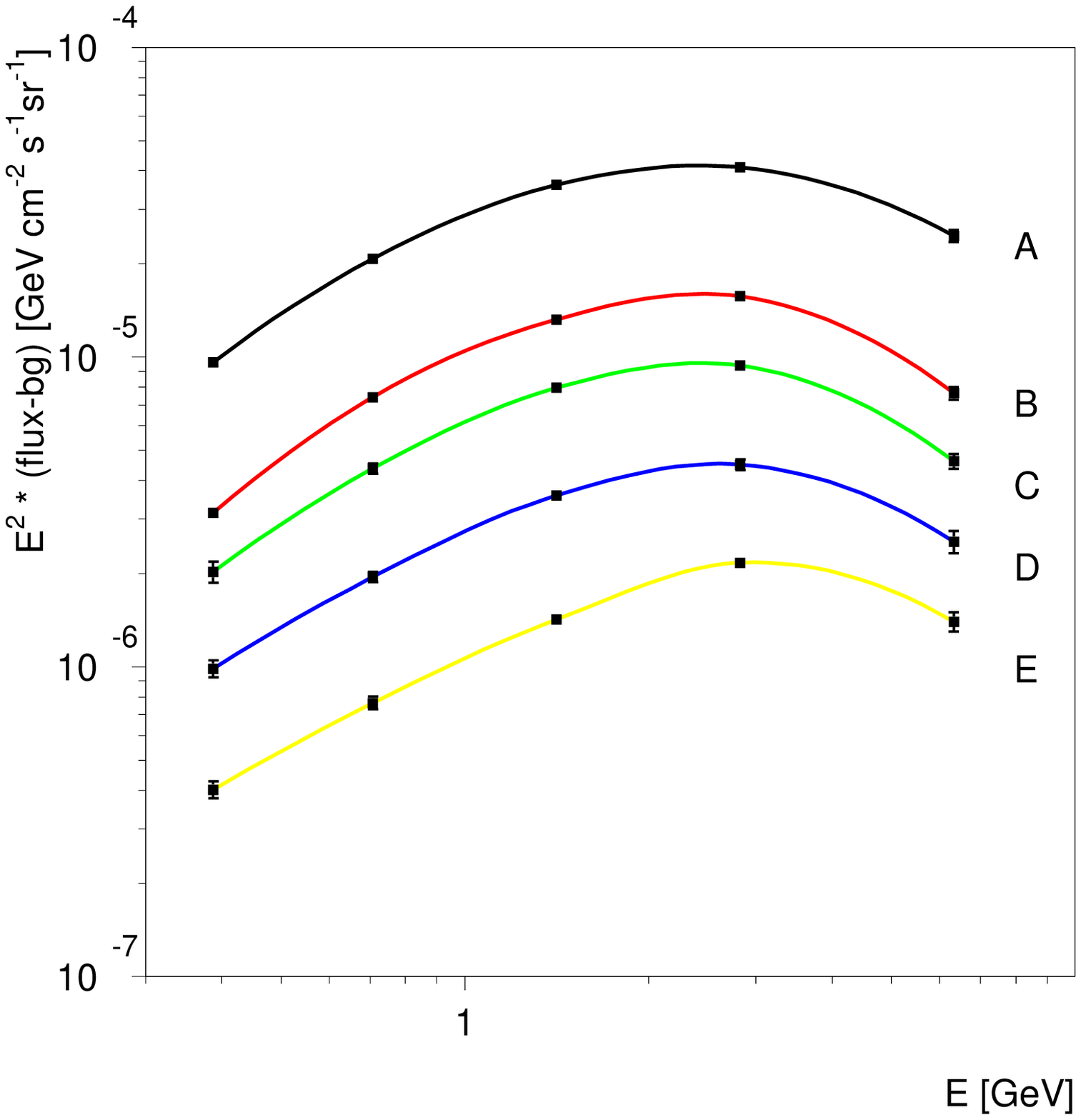}
\includegraphics [width=0.45\textwidth,clip]{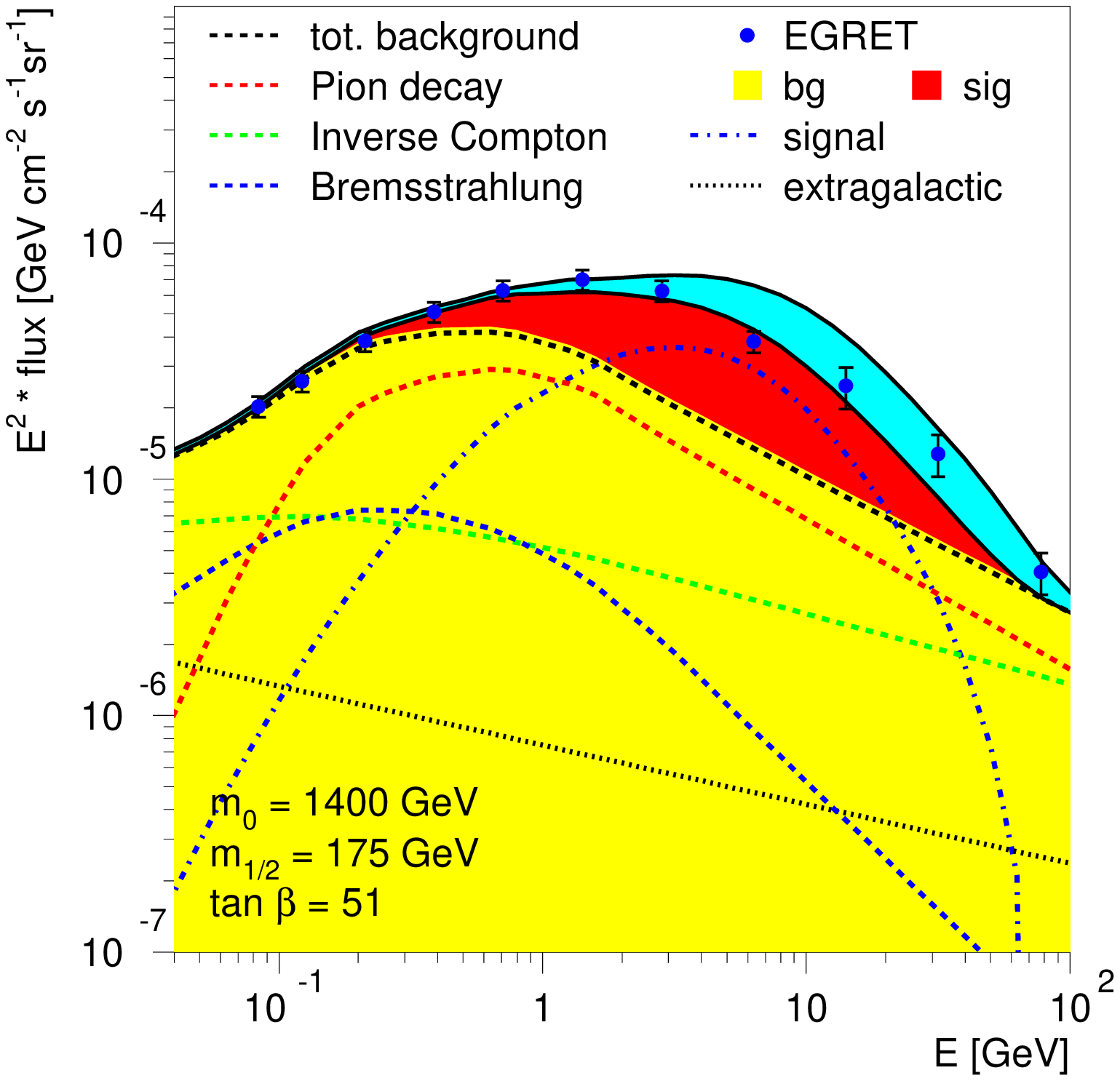}
 \caption[]{
Left: the energy spectrum of the EGRET excess, defined as the
difference between the data and the fitted background contribution
for 5 of the panels of Fig. \ref{excess}.
Only the small statistical errors have been plotted.
Right: the medium shaded (blue) band shows the effect of varying the WIMP mass
between 65 and 100 GeV.}
 \label{diff}
\end{center}
\end{figure}

\begin{figure}[t]
\begin{center}
 \includegraphics [width=0.35\textwidth,clip]{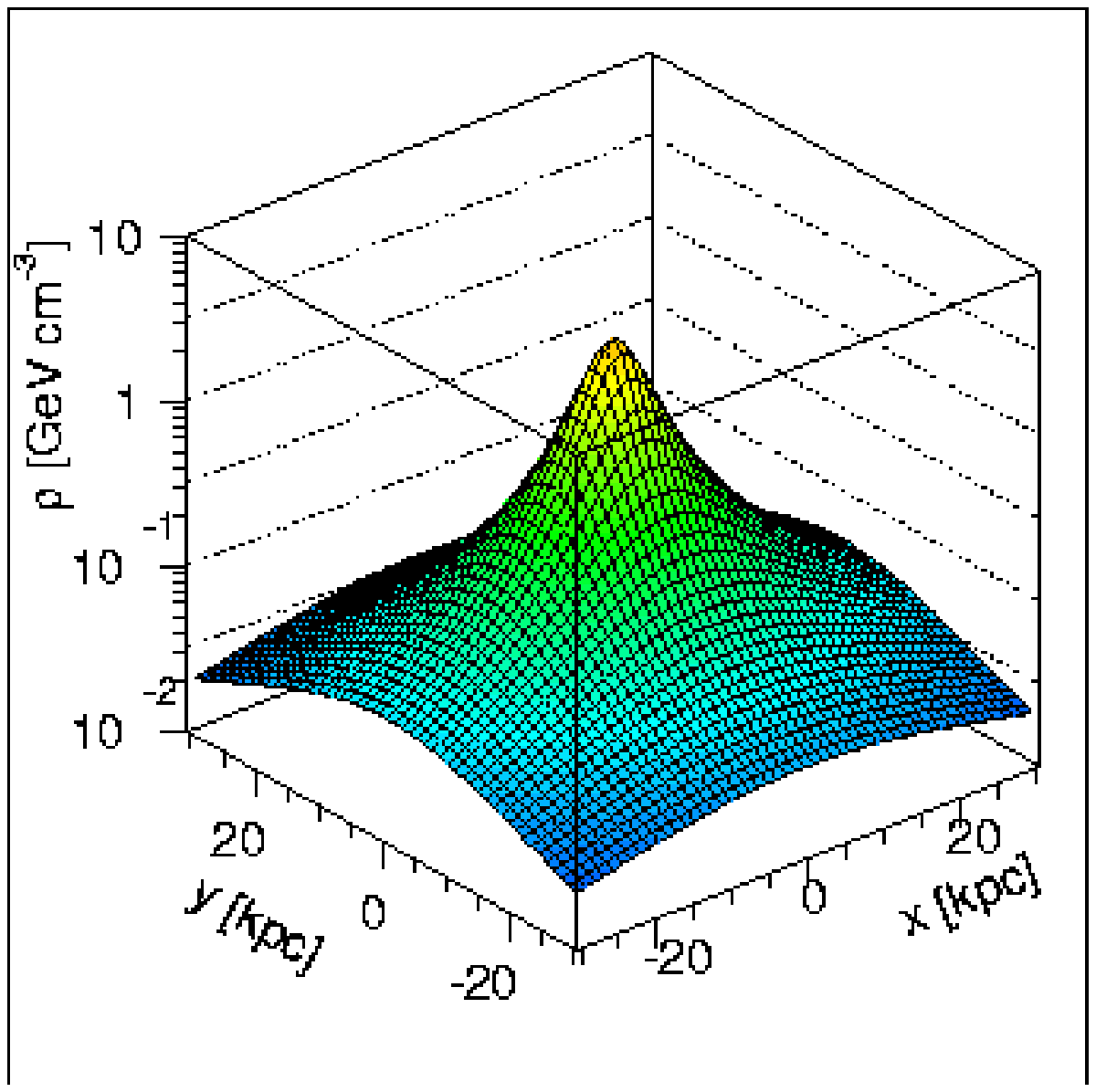}
\includegraphics [width=0.35\textwidth,clip]{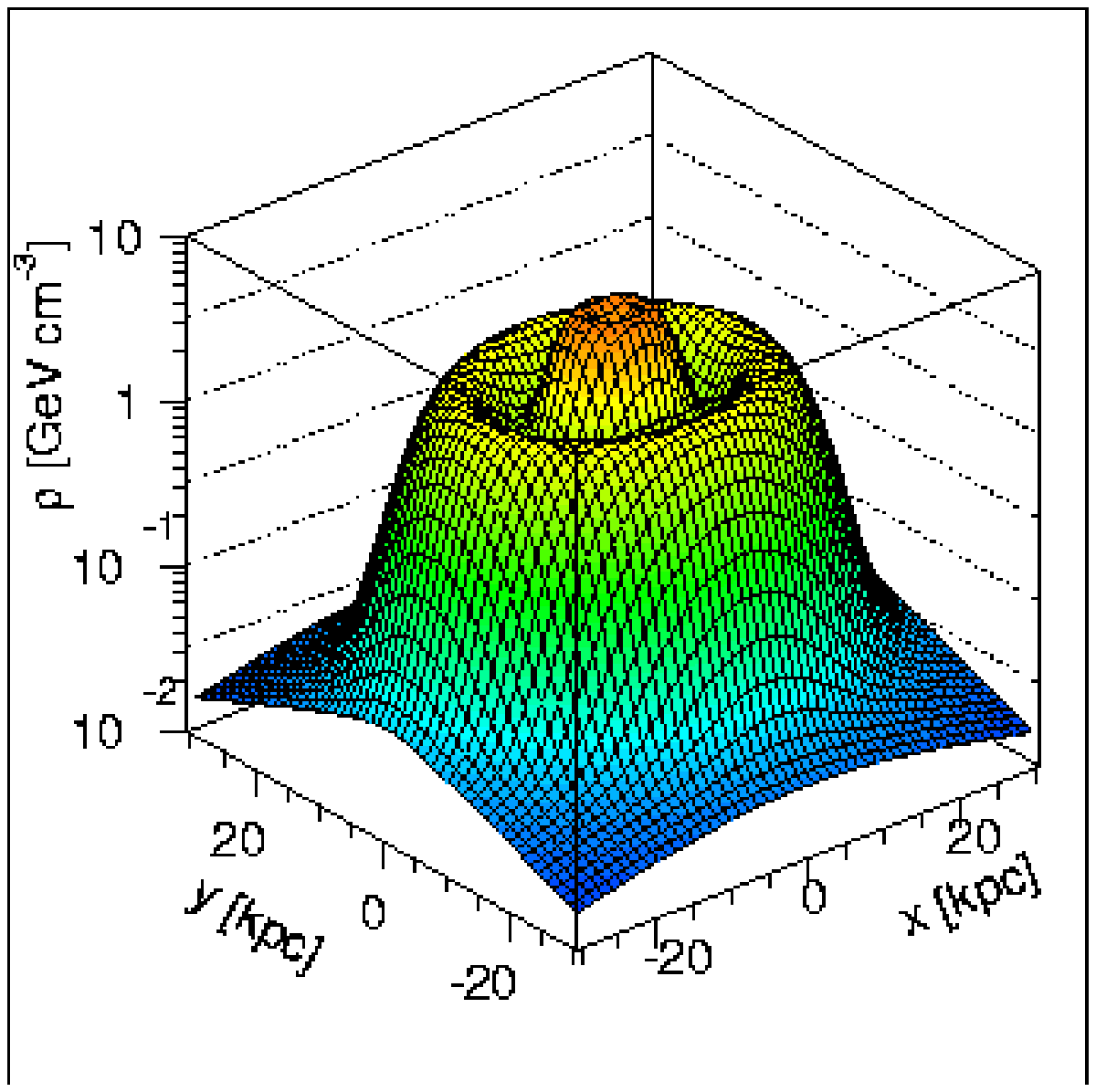}
\includegraphics [width=0.35\textwidth,clip]{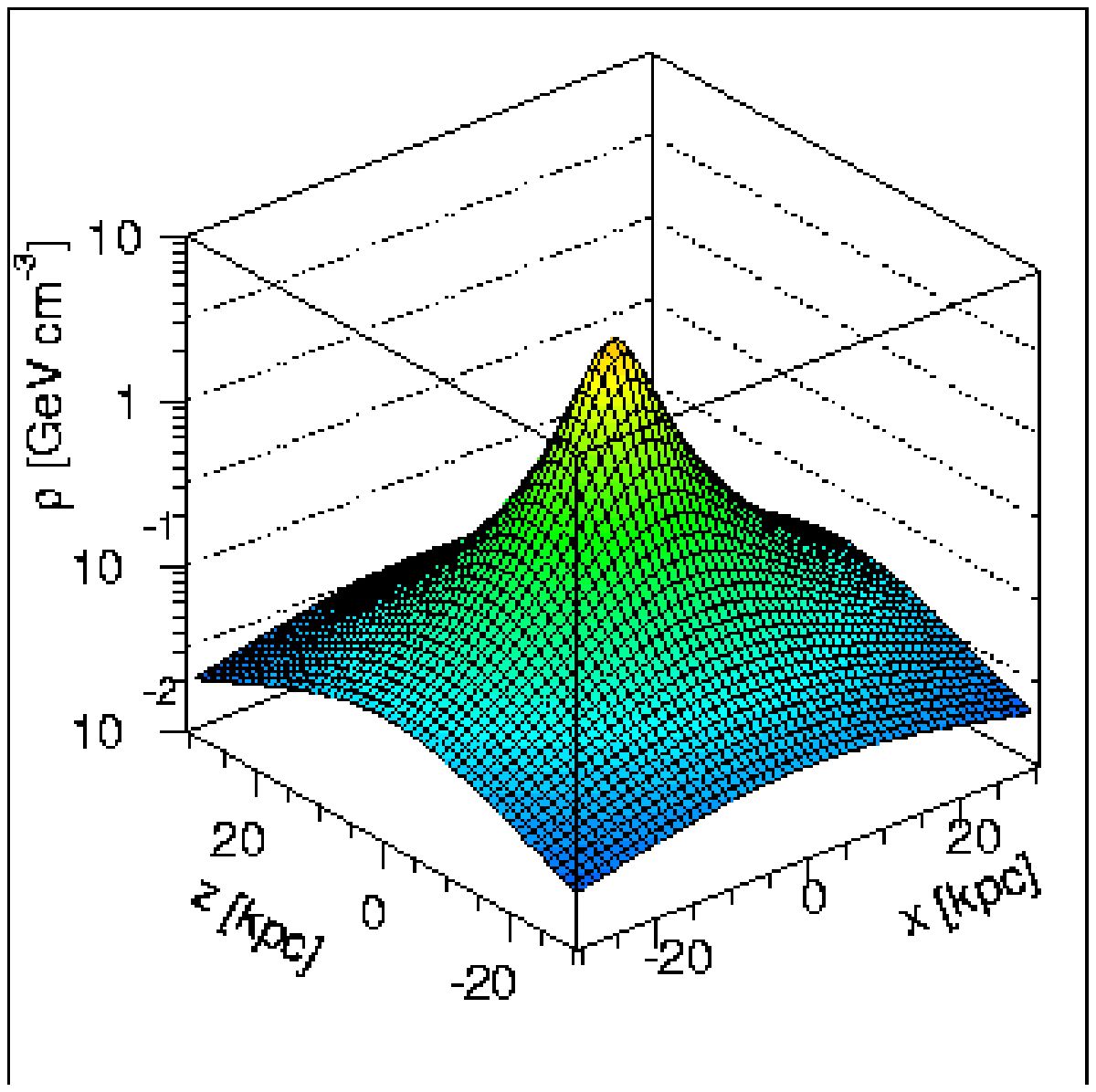}
\includegraphics [width=0.35\textwidth,clip]{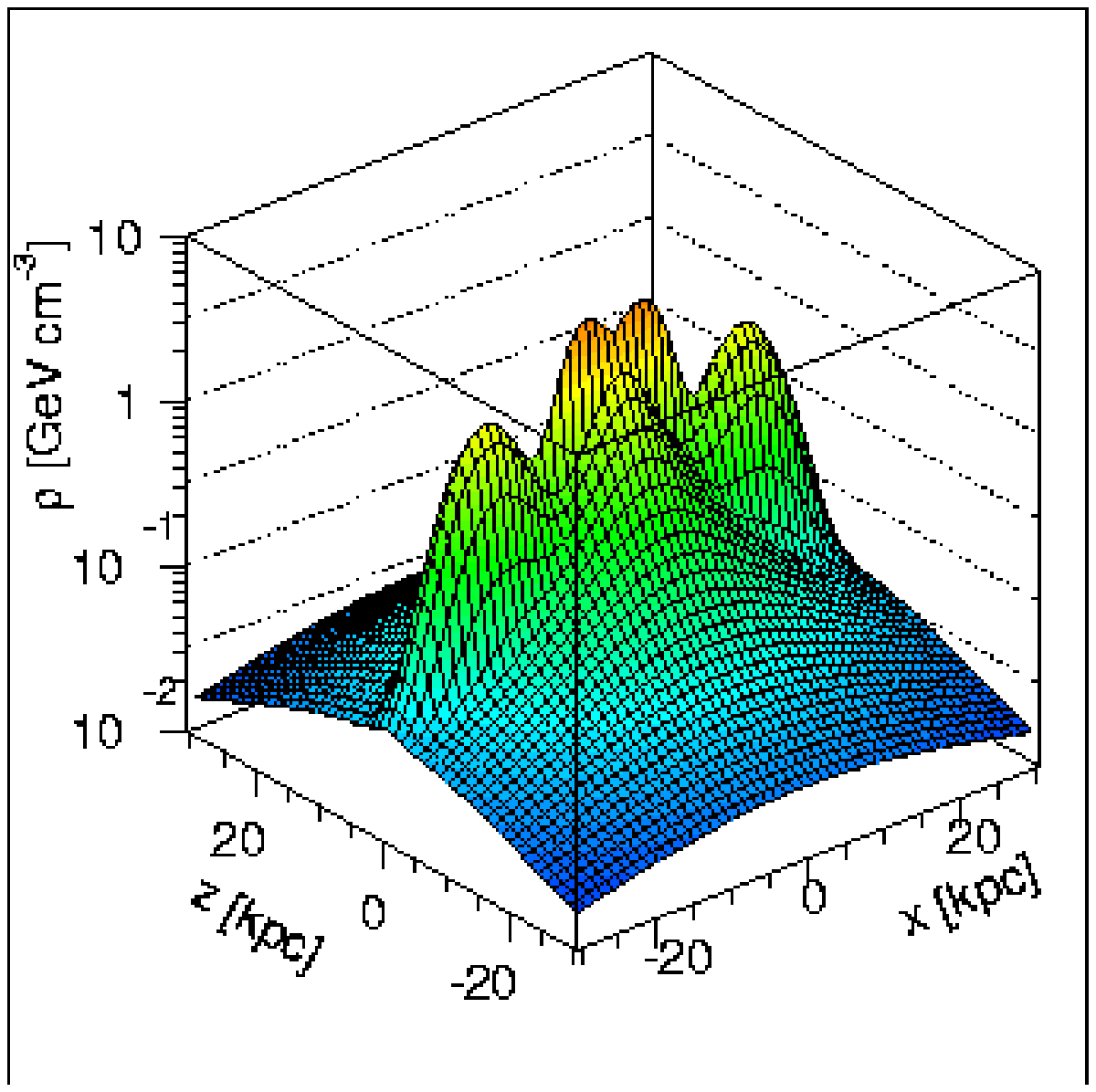}
 \caption[]{
  3D-distributions of the $1/r^2$ haloprofile
  in the galactic xy-plane  (top row) and xz-plane (bottom row)
  without (left) and with (right) rings. In the disc (xy-plane) the vertical
axis shows clearly the enhancement in density at a radius of 14 kpc.
Going along the x-axis  below the disc (negative z) shows  the $1/r^2$ profile in the xz-plane,
while going along the x-axis in the disc (z=0) shows the ring structure.
}
 \label{profile}
\end{center}
\end{figure}

 A  summary of these discussions have been given by Strong et
al.\cite{optimized}, who find that hard proton spectra are
incompatible with the antiproton yield and hard electron spectra are
incompatible with the EGRET data up to 120 GeV, which they analyzed.
However, they find that by modifying the electron and proton
injection spectra simultaneously, they can improve the description
of the data, as noted also recently by Kamae et al.\cite{kamae}.

The problem with these "solutions" is that they give a too large
(small) contribution at low (high) gamma ray energies, i.e. the
shape of the energy spectra is not well reproduced. But it is
exactly the shape, which was well measured by EGRET, because the
quoted normalization errors of 15\% are common to all energy points.
If one calculates the probability of the "optimized" model, taking
the correlations between the energy points into account, the
probability is below 10$^{-14}$! Two other arguments, independent of
the EGRET errors, against "optimized" models are: 1) the energy loss
time of protons above 10 GeV is above $10^{11}$ yrs, i.e. longer
than the lifetime of the universe. Therefore it is hard to image
that protons, accelerated in the centre of the galaxy by the many
supernovae there, would have a significant different spectrum after
diffusion to the solar neighbourhood in about $10^8$ yrs, a time
much shorter than the energy loss time 2)
 if the proton spectrum is nevertheless inhomogeneous over our
 galaxy, it is very surprising that the excess has the {\it same}
 energy shape towards the outer galaxy, where there are practically no supernovae
  and towards the centre of the galaxy.
An alternative way of formulating this problem: if the EGRET excess
can be explained prefectly in all sky directions by a gamma
contribution originating from  mono-energetic quarks, it is very
difficult to replace such a contribution by an excess  of quarks (or
electrons) with a power law spectrum.

To exemplify these problems we consider the {\it shape} of the
background from a recent  analysis by Kamae et al.\cite{kamae}. They
use a harder proton spectrum than locally observed (a power law with
index 2.5 instead of 2.7 observed locally) and an updated pp cross
section including diffractive scattering and scaling violation. They
claim this can describe the EGRET data towards the galactic centre.
However, there is a clear overshoot at low energies. Fitting only
their shape to the EGRET data, i.e. with a free normalization, still
leaves a significant excess, as shown in the left panel of Fig.
\ref{excess}. Here the upper  edge of the medium shaded (blue) area
corresponds the hardest possible spectrum
 from Kamae et al.\cite{kamae} with the power index of 2.5, while
the lower edge corresponds to the conventional GALPROP
model\cite{optimized}. Note that the hard spectrum overshoots the
highest EGRET point, which was not yet available during the analysis
by Kamae et al. In summary, also for the ``conventional''
explanations\cite{optimized, kamae} the fit to {\it all} sky
directions can be much improved, if DM is added, since then both the
low and high energy range can be perfectly described. Thus different
backgrounds just change the normalization of the DM contribution.

\begin{figure}[t]
\begin{center}
 \includegraphics [width=0.36\textwidth,clip]{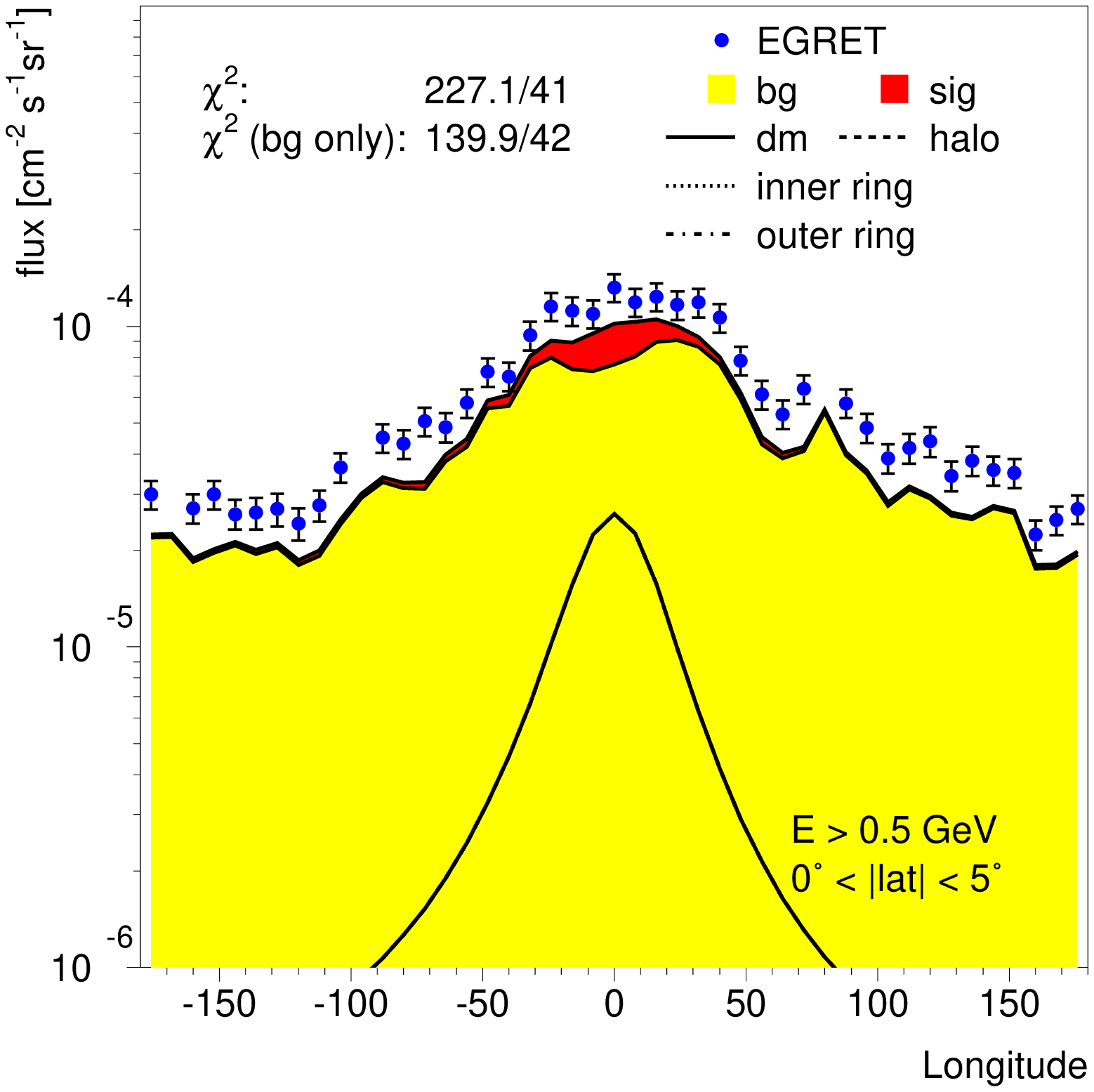}
 \includegraphics [width=0.36\textwidth,clip]{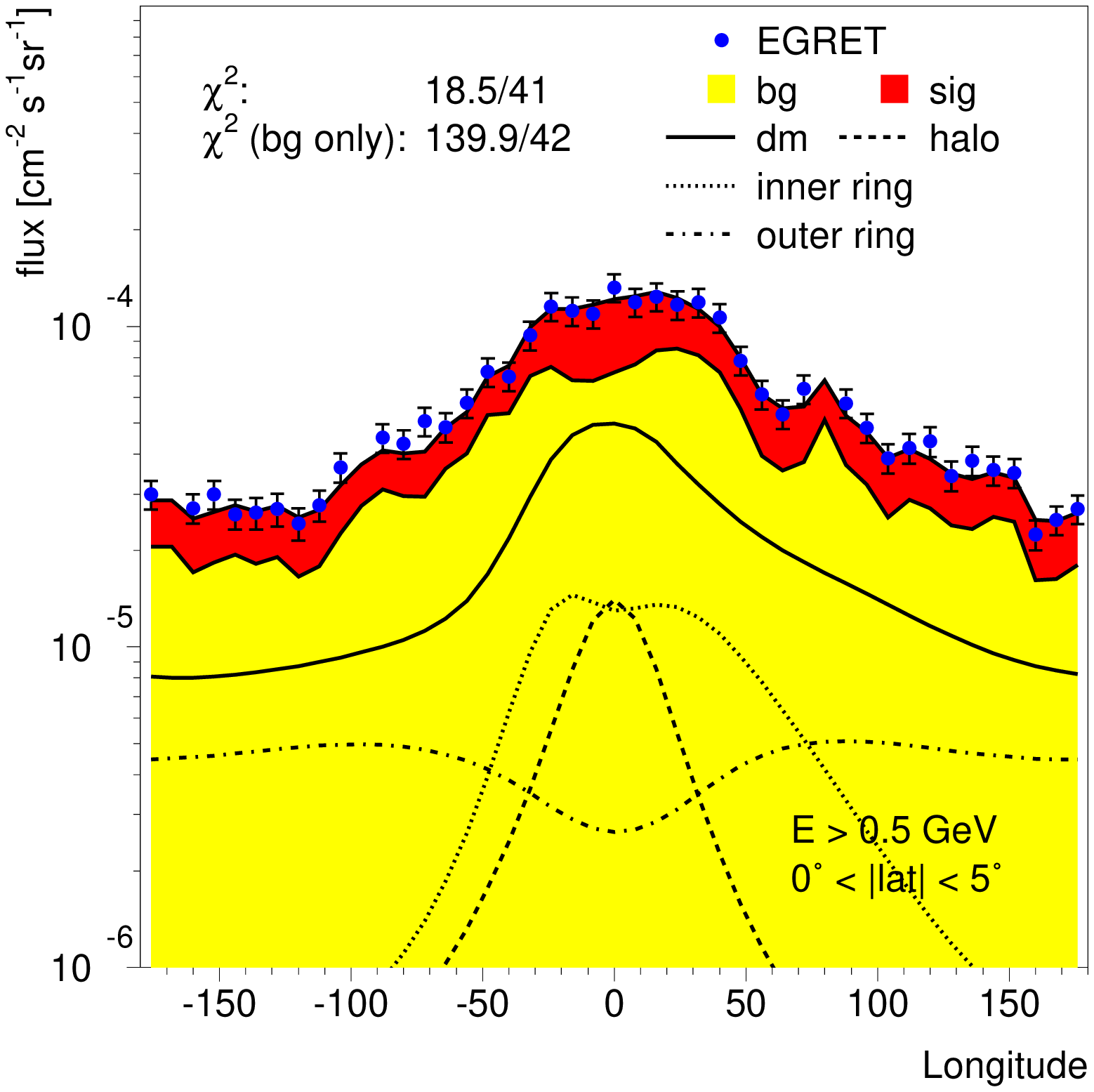}
 \includegraphics [width=0.36\textwidth,clip]{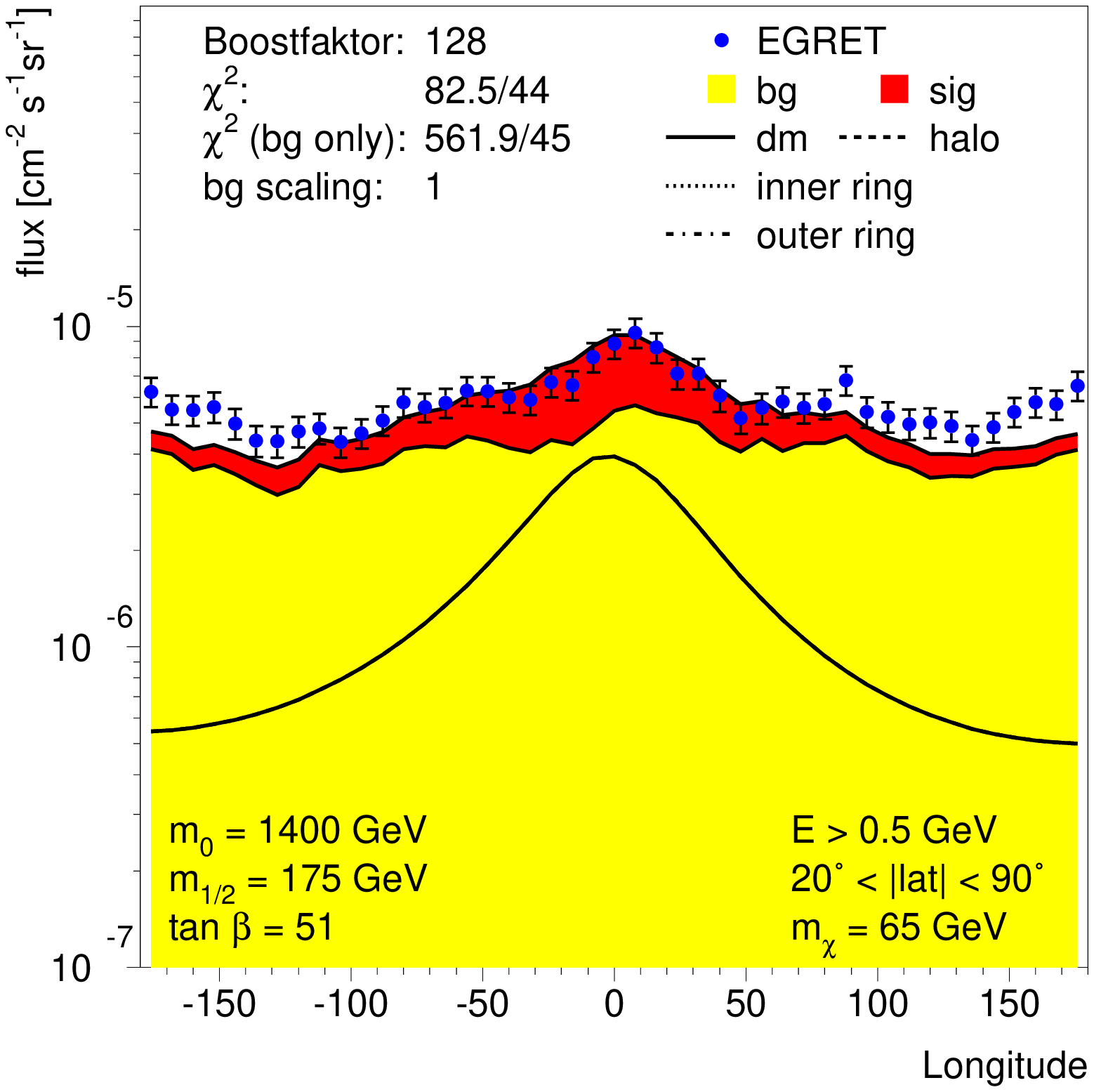}
 \includegraphics [width=0.36\textwidth,clip]{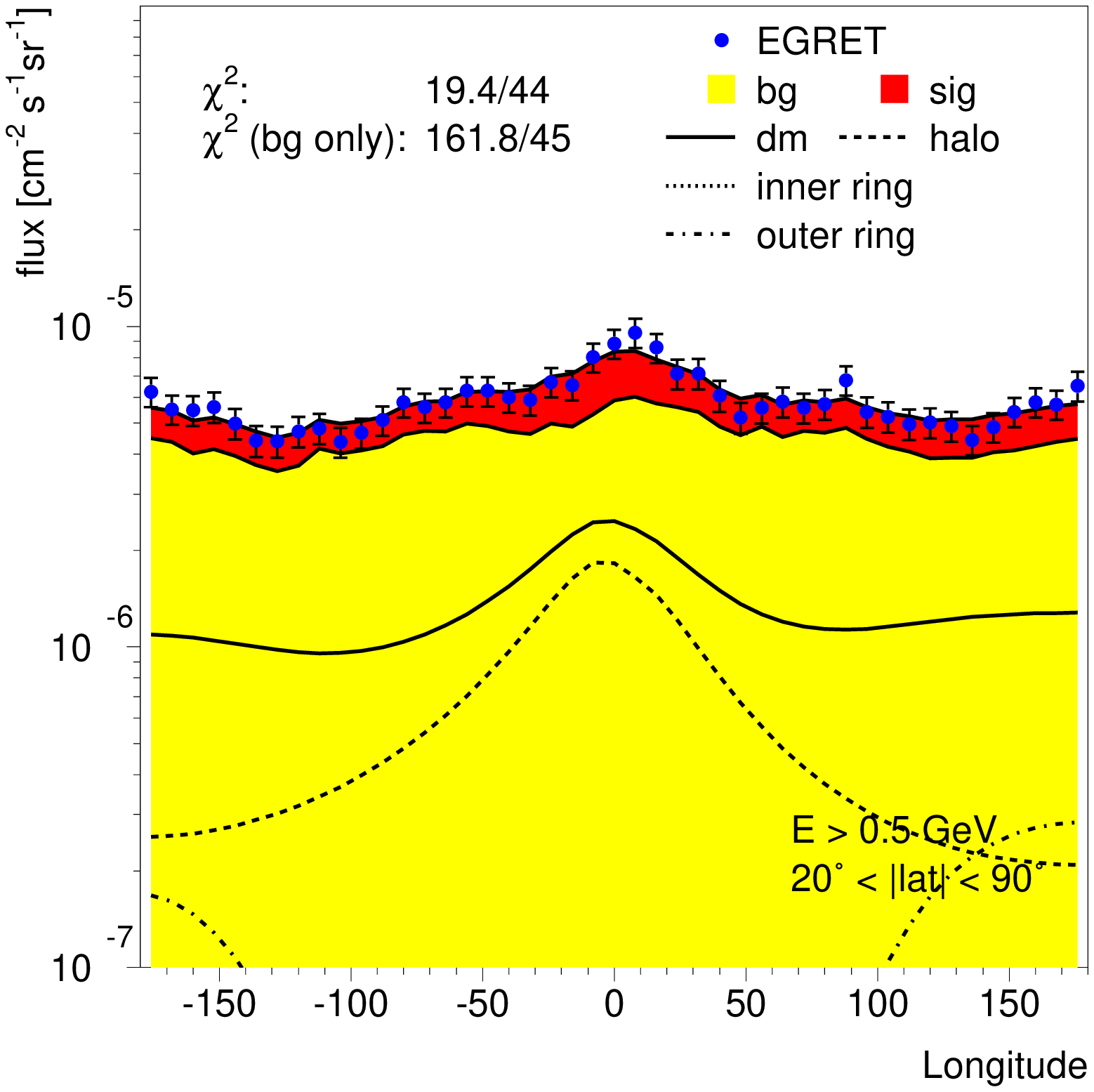}
 \caption[]{Top row:
 the longitude distribution of diffuse gamma-rays in the disc of the galaxy
 (latitudes $0^\circ<|b|<5^\circ$)
for the $1/r^2$ profile without (left) and with rings (right).
 The points represent the EGRET data.
  Bottow row: as above for the polar regions of our galaxy (latitudes  $20^\circ<|b|<90^\circ$) .}
 \label{long}
\end{center}
\end{figure}

The quality of the EGRET data is better appreciated if one plots
only the statistical errors. Fig. \ref{diff} shows the excess for
five different sky regions: only at high latitudes the errors start
to be visible. The curves are just spline fits through the data and
were used to determine the systematic point-to-point errors by
leaving a given energy point out of the fit and determine its
variance. The point-to-point error is about 7\% for most energy
points. In the previous plot the WIMP mass was kept constant at 65
GeV. The right hand side of Fig. \ref{diff} shows the plot for a
WIMP mass of 100 GeV, which clearly overshoots the high energy data.
Therefore a rough estimate of the WIMP mass from the EGRET excess is
between 50 and 100 GeV.

\begin{figure}[t]
\begin{center}
 \includegraphics [width=0.6\textwidth,clip]{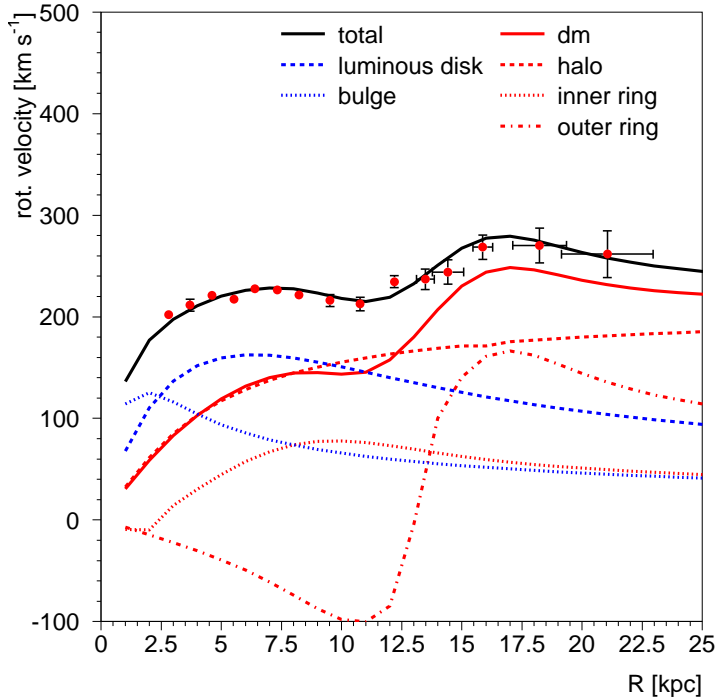}
 \caption[]{
  The rotation curve from our galaxy with the DM contribution determined from the
   EGRET excess of diffuse gamma rays. The data are averaged from Ref. \cite{deboer}.}
 \label{rot}
\end{center}
\end{figure}

From the excess in the various sky directions one can obtain the
halo profile under the assumption that the clustering of the DM is
similar in all sky directions. The result is surprising: in addition
to the $1/r^2$ profile expected for a flat rotation curve the EGRET
excess show a  substructure in the form of toroidal rings at 4 and
14 kpc, as shown in Fig. \ref{profile}: on the left hand side the
contribution from the $1/r^2$ profile is shown, while for the right
hand side the ring structure is added. The need for these additional
rings is most easily seen by comparing the longitudinal profiles in
the galactic plane and towards the galactic poles. As shown in Fig.
\ref{long} the pole regions are described reasonably well without
rings, but for the galactic plane the $1/r^2$ profile only describes
the data towards the centre. For the larger latitudes one needs the
rings, as indicated by the right top panel. Note that for each bin
only the flux integrated for data above 0.5 GeV has been plotted.
The normalization of the background has been obtained from a fit to
the flux integrated between 0.1 and 0.5 GeV.

 The position and shape of the inner ring coincides
with the ring of molecular hydrogen. Molecules form from atomic
hydrogen in the presence of dust or heavy nuclei. So a ring of
neutral hydrogen suggests an attractive gravitational potential. The
position and shape of the outer ring coincides with the ring of
stars, discovered in 2003 by two independent
groups\cite{yanny,ibata}. This ring is thought to originate from the
infall of a dwarf galaxy, so additional DMA is expected there.

To prove that the enhanced gamma ray density is indeed connected to
non-baryonic mass the rotation curve was reconstructed from the
excess of the diffuse gamma rays in the following way: since the
flux determines the number density of DM for a given boost factor
and since the mass of each WIMP is between 50 and 100 GeV, one can
determine the mass in the ring and consequently predict the rotation
curve\footnote{For the outer ring a total DM mass of a few times
$\rm 10^{10}$ solar masses was found in comparison with about $\rm
10^9$ solar masses in the form of stars.}. The two ring model
describes the peculiar change of slope at 11 kpc well, as shown in
Fig. \ref{rot}. The contributions from each of the mass terms have
been
 shown separately. The basic explanation for the negative contribution from the outer ring
 is that a tracer star at the
 inside of the ring at 14 kpc feels an outward force from the ring, thus a negative
 contribution to the rotation velocity.
 It has often been argued that  the outer
 rotation curve cannot be taken seriously, because
the errors are large due to the fact that the absolute values of the
rotation velocities strongly depend on the value of $R_0$, the
distance between the solar system and the galactic centre. This is
true, as shown by Honma and Sofue\cite{honma}, but they show that
the {\it change in slope} at about 1.3$R_0$  is independent of
$R_0$. In addition, it has been argued that the inner and outer
rotation curve are difficult to compare, since the methods are
completely different. The methods are indeed different, but
 the first 3 data points from the outer rotation curve
  (between 8 and 11 kpc) show the same slope as the ones from the
  inner rotation curve, so there seems to be no systematic effect
  related to the different methods.
\begin{figure}
\begin{center}
 \includegraphics [width=0.45\textwidth,clip]{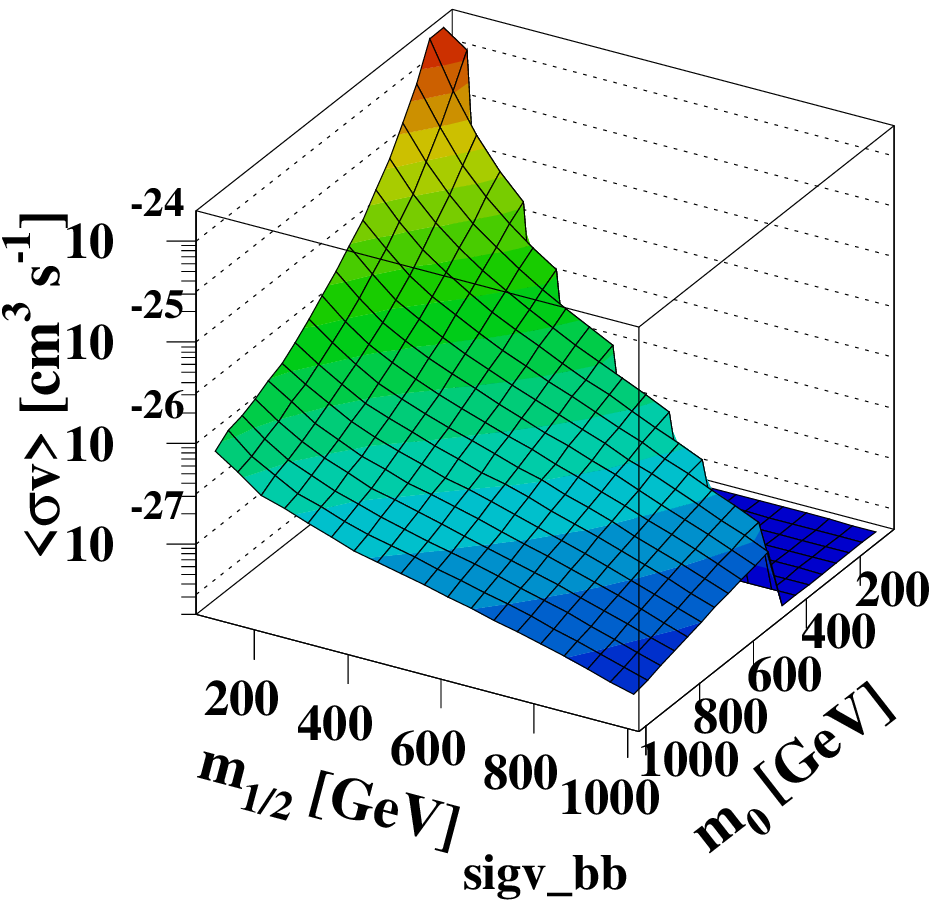}
 \includegraphics [width=0.45\textwidth,clip]{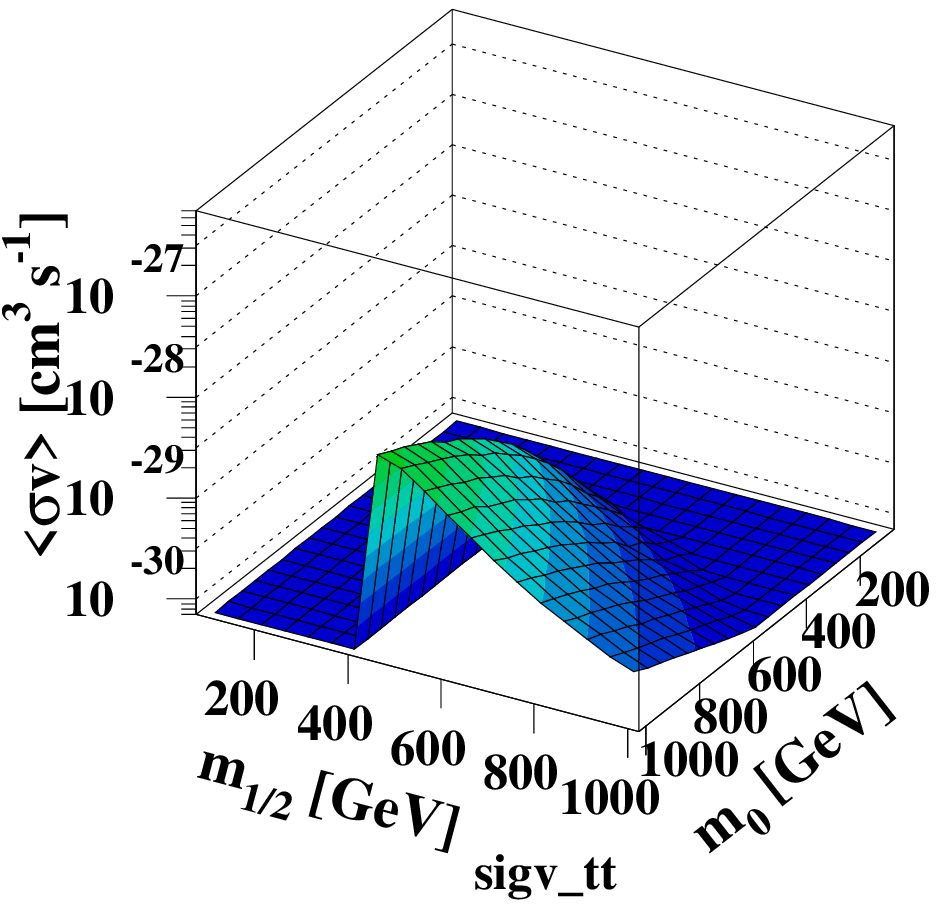}
 \includegraphics [width=0.45\textwidth,clip]{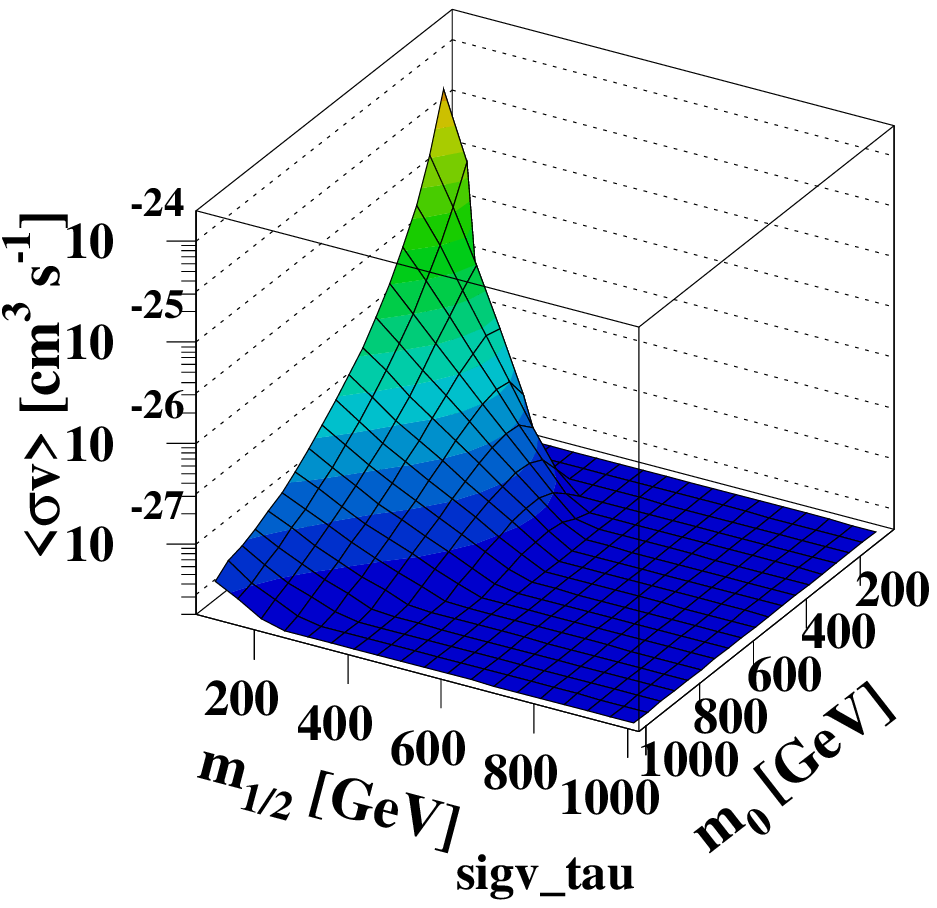}
 \includegraphics [width=0.45\textwidth,clip]{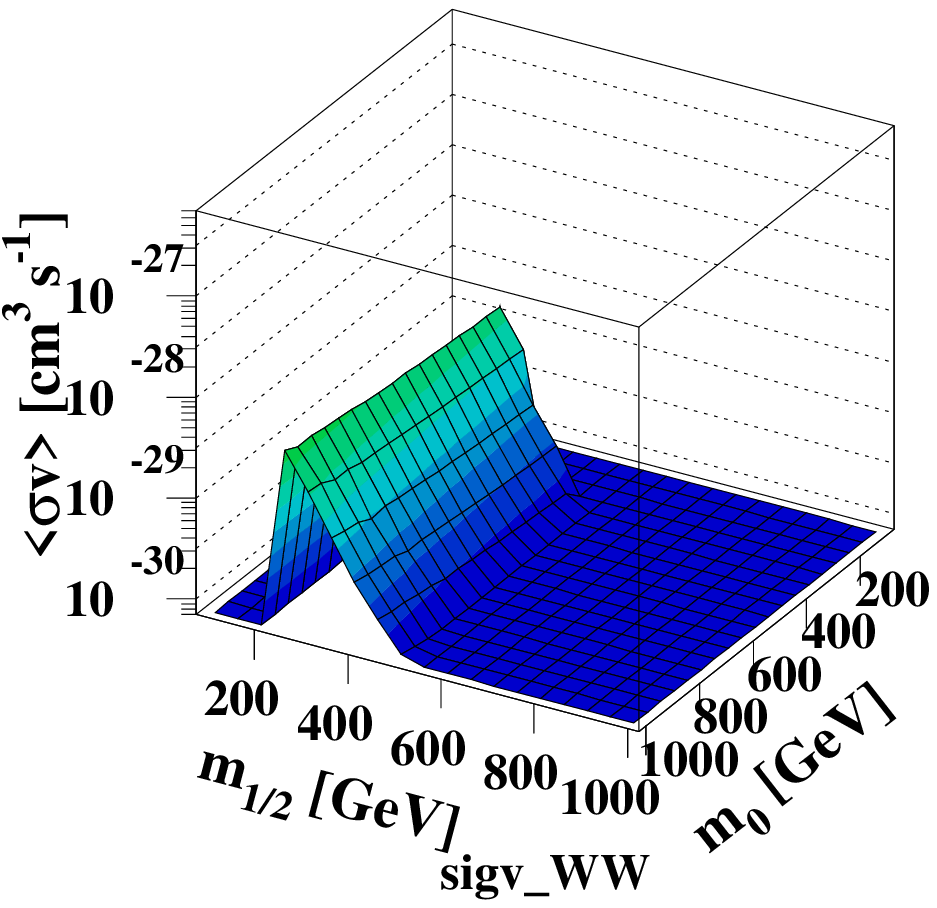}
 \includegraphics [width=0.45\textwidth,clip]{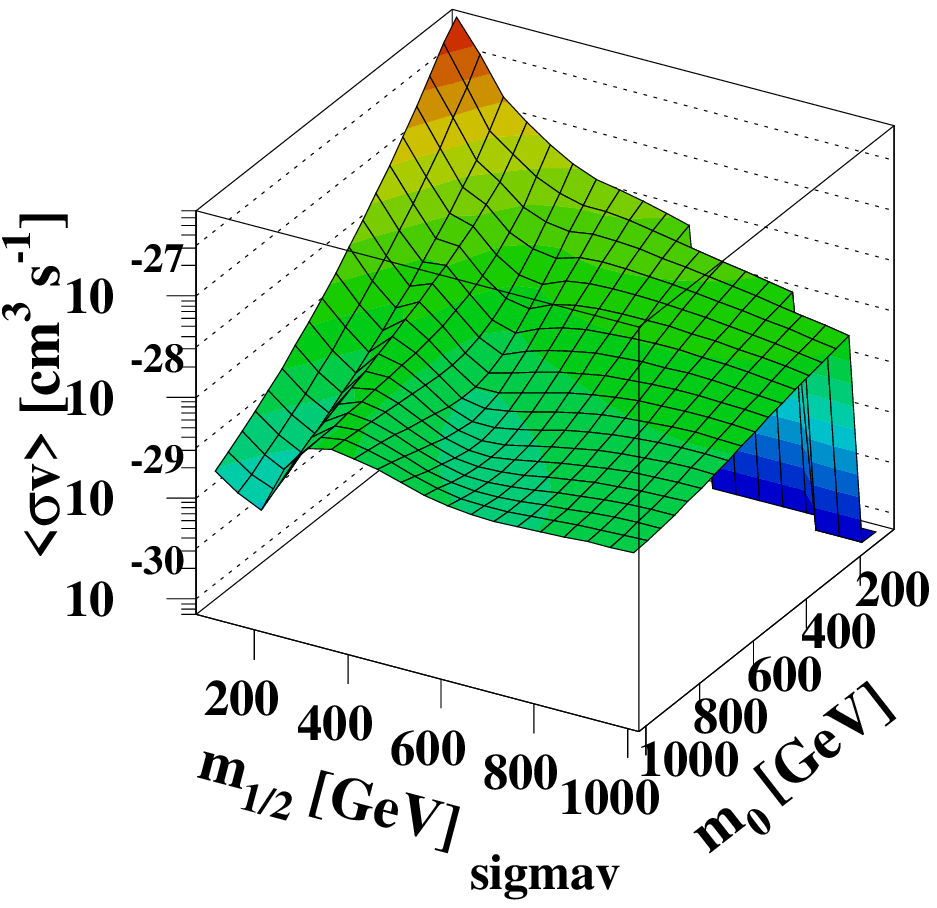}
 \includegraphics [width=0.45\textwidth,clip]{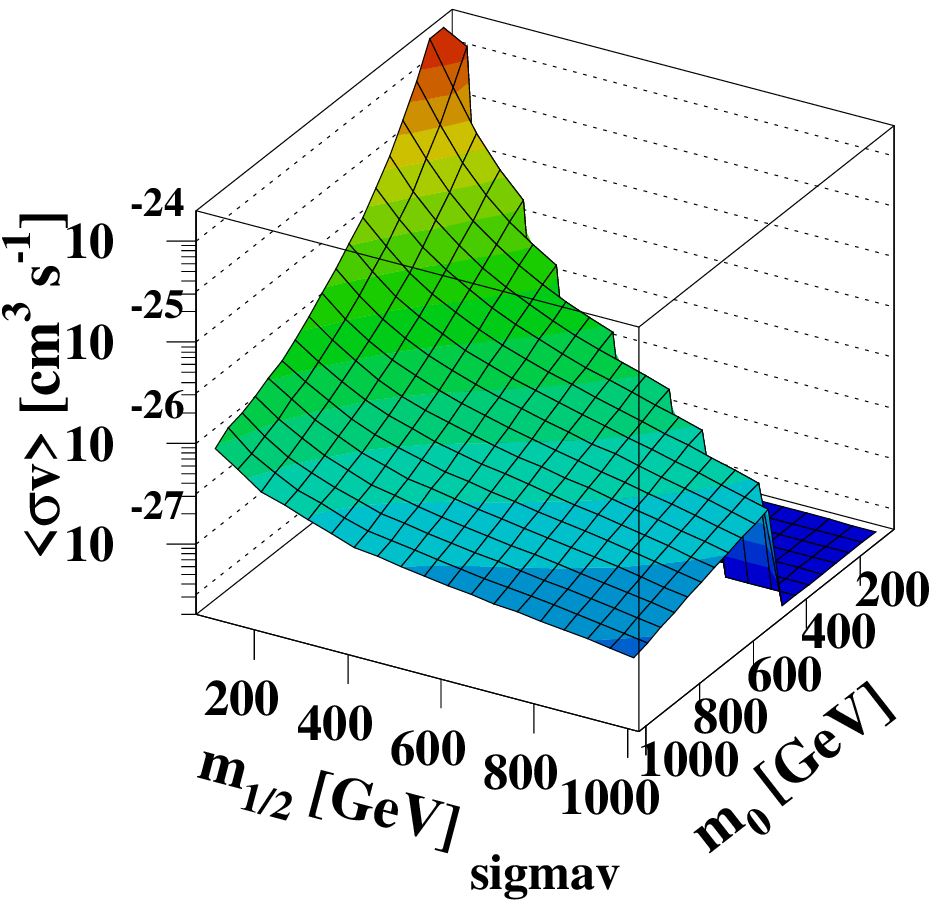}
 \caption[]{\label{sigmavall} The first two rows show
 the thermally averaged annihilation cross section times velocity for
 neutralino annihilation as function of $m_0$ and $m_{1/2}$ for
  $\tan\beta$= 50 and $b\overline{b}$, $t\overline{t}$,
  $W^+W^-$, and $\tau\overline{\tau}$ final states (clockwise from top left).
  The last row shows the total cross section or
 $\tan\beta$= 5 (left) and 50 (right). The neutralino mass equals $\approx 0.4 m_{1/2}$
 in the CMSSM, so the neutralino varies from 40 to 400 GeV along the
 front axis. Note the strong decrease of the cross section for
 heavier SUSY mass scales and the different vertical scales.
 }
\end{center}
\end{figure}

\begin{figure}
\begin{center}
 \includegraphics [width=0.45\textwidth,clip]{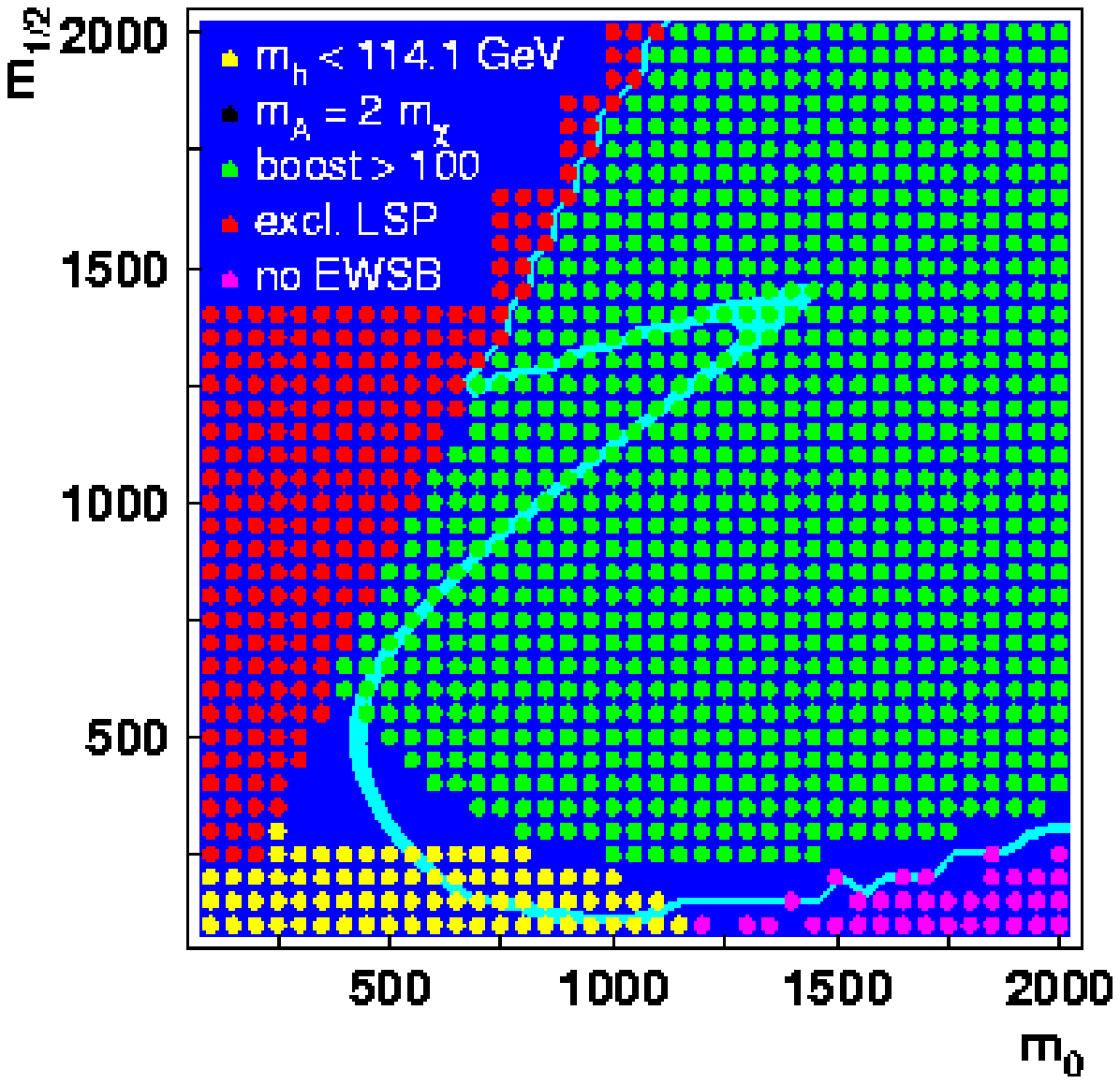}
 \includegraphics [width=0.45\textwidth,clip]{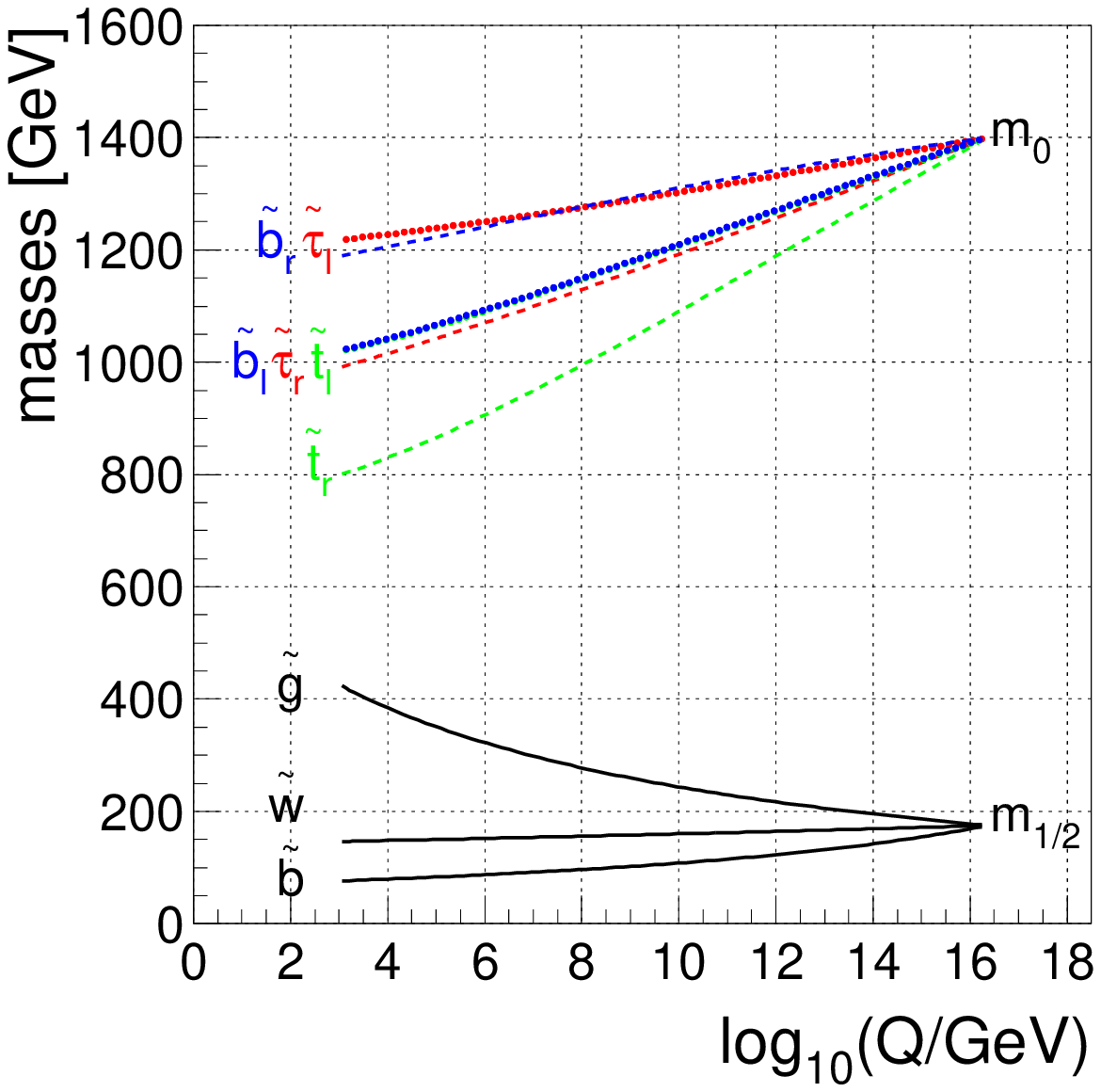}
 \caption[]{
 The light shaded (blue) line in the region allowed by WMAP in the $\mzero,\mhalf$ plane
for $\tb=51$, $\mu>0$ and $A_0=0$.
  The excluded regions, where the stau would be the LSP or EWSB fails or
 are indicated by the dots. The large central region, where the boost factor would be above
 100, has been indicated as well.
The region for $\mhalf\approx 180$ and $\mzero\approx 1400$ all
constraints from EGRET, WMAP and electroweak data are fulfilled. The
evolution of the  particle spectrum from the GUT scale is shown on the right hand side,
showing that the squarks and sleptons have masses in the TeV range,
while the gluinos and charginos are relatively light.
 }
 \label{relic}
\end{center}
\end{figure}

\section{Comparison with Supersymmetry}

Supersymmetry~\cite{susyrev} presupposes a symmetry between fermions
and bosons, which can be realized in nature only if one assumes each
particle with spin j has a supersymmetric partner with spin $\vert
j-1/2\vert$ ($\vert j-1/2\vert$ for the Higgs bosons). This leads to
a doubling of the particle spectrum. Obviously SUSY cannot be an
exact symmetry of nature; or else the supersymmetric partners would
have the same mass as the normal particles. The mSUGRA model, i.e.
the Minimal Supersymmetric Standard Model (MSSM) with supergravity
inspired breaking terms, is characterized by only 5 parameters:
$m_0,~m_{1/2},~\tb,~\mbox{sign}(\mu), ~A_0$. Here $m_0$ and
$m_{1/2}$ are the common masses for the gauginos and scalars at the
GUT scale, which is determined by the unification of the gauge
couplings. Gauge unification is still possible with the precisely
measured couplings at LEP~\cite{bs}. The ratio of the vacuum
expectation values of the two Higgs doublets is called \tb ~ and
$A_0$ is the trilinear coupling at the GUT scale. We only consider
the dominant trilinear couplings of the third generation of quarks
and leptons and assume also $A_0$ to be unified at the GUT scale.
The absolute value of the Higgs mixing parameter $\mu$ is determined
by electroweak symmetry breaking, while its sign is taken to be
positive, as preferred by the anomalous magnetic moment of the
muon~\cite{bs}.
\begin{figure}
\begin{center}
 \includegraphics [width=0.45\textwidth,clip]{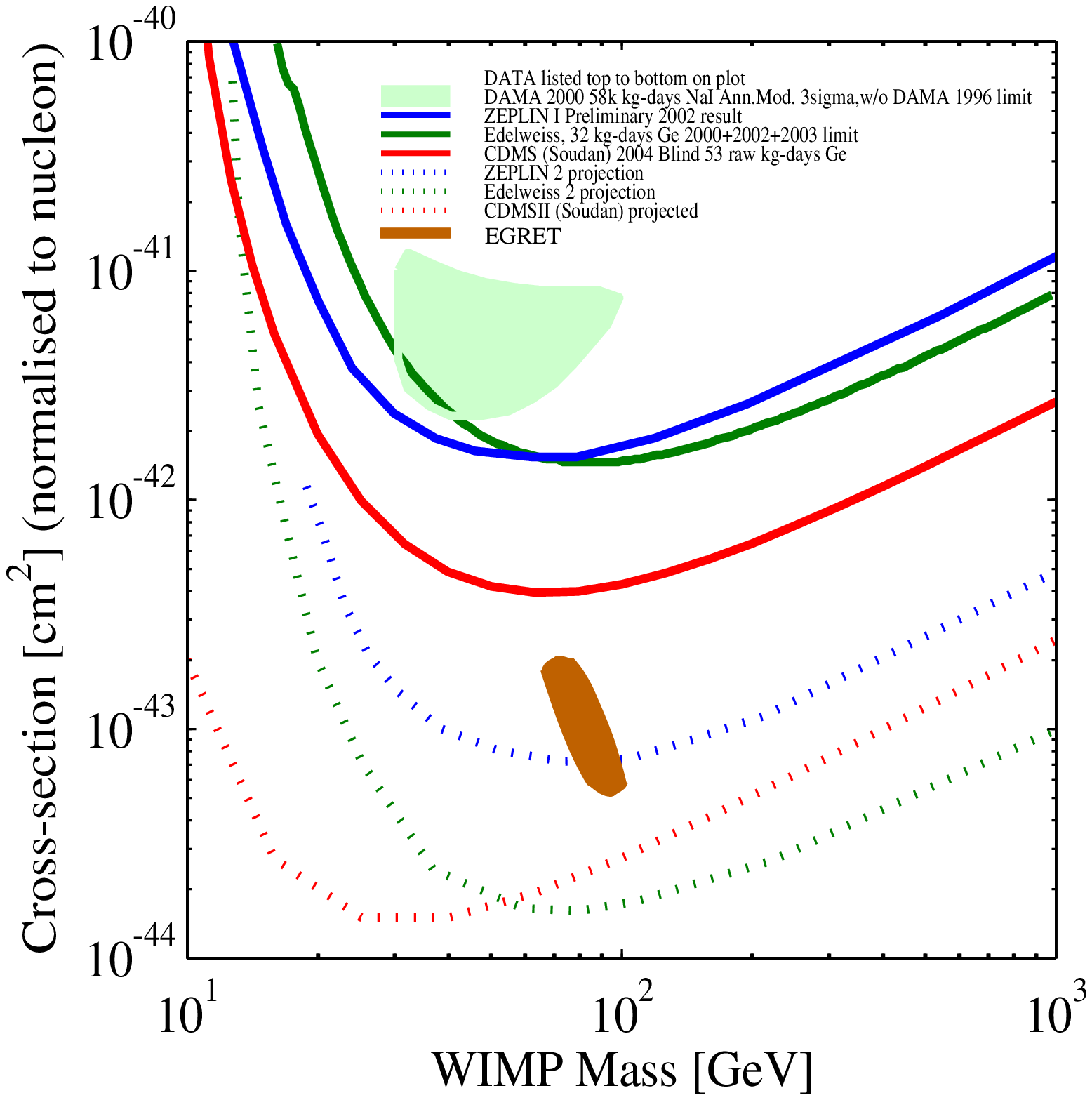}
\includegraphics [width=0.45\textwidth,clip]{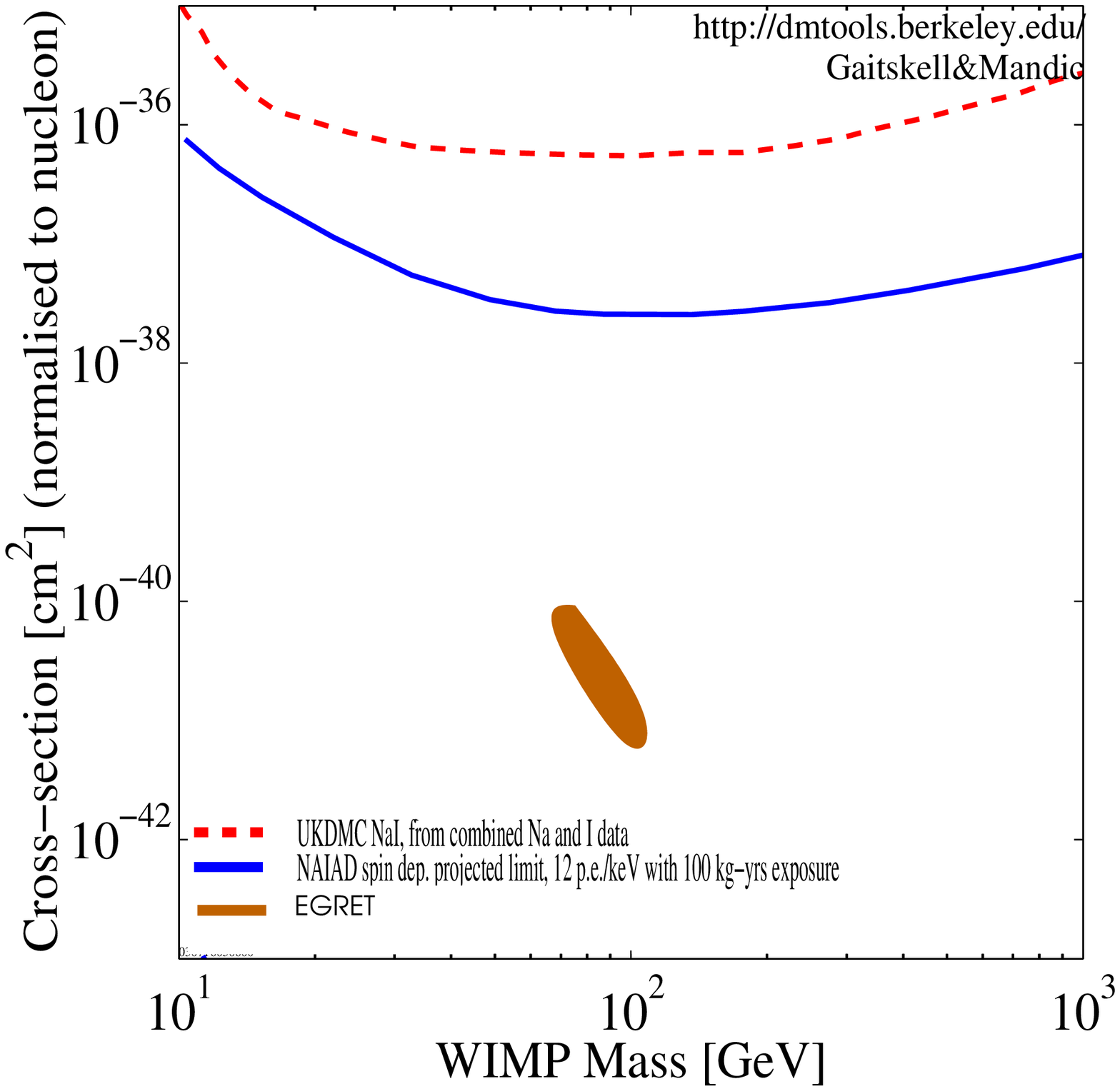}
 \caption[]{
 The spin-independent (left) and spin-dependent (right)
 neutralino-nucleon cross section  as function of the neutralino
 mass for the SUSY parameters from this analysis\cite{deboer} (oval
 shaded (brown) area in comparison with results from present and future direct DM
 detection experiments.
 }
 \label{direct}
\end{center}
\end{figure}
The lightest supersymmetric particle (LSP) is stable, if the
multiplicative quantum number R-parity, which is +1 for SM particles
and -1 for SUSY particles, is conserved. Non-conservation of
R-parity would lead to rapid proton decay\cite{susyrev}. The LSP is
a perfect candidate for Dark Matter and it can self annihilate into
fermion-antifermion pairs by Higgs or Z-exchange in the s-channel or
sfermion, chargino and neutralino exchange in the t-channel. The
dominant first three possibilities have amplitudes proportional to
the fermion mass, so heavy final states are preferred. For  values
of $\tb\approx 50$ the annihilation cross sections into
$b\overline{b}$ quarks are indeed of the order of magnitude required
by WMAP, as shown in Fig. \ref{sigmavall}. For $m_{1/2}\approx 175$
GeV, corresponding to a neutralino mass of about 70 GeV, as required
by the EGRET data, the scalar masses have to be in the TeV range for
a thermally averaged annihilation cross section $\sigma v\approx 2.
10^{-26} ~cm^3/s$, as required by Eq. \ref{wmap}.

This point of parameter space respects in addition all constraints
from the direct searches on Higgs and charginos and electroweak
precision observables, as shown in Fig. \ref{relic}. The relic
density has been calculated with the program MicroMegas\cite{micro}.
If $m_0$ is small compared with $m_{1/2}$ the lightest lepton
(usually the stau) can be lighter than the neutralino, which happens
in the left top corner on the left hand side of Fig. \ref{relic}. In
the region adjacent to it the stau cannot decay fast into a
neutralino and tau,  in which case a stau and neutralino can
annihilate into a tau plus photon. This coannihilation reduces the
relic density to values required by the WMAP data, but these regions
require large boost factors, since in the present galaxy the NLSP's
have decayed and only the self annihilation contributes. The
regions, where the boost factors are above 100 are shown in Fig.
\ref{relic} together with the regions where the annihilation cross
section is consistent with the WMAP data. For boost factors below
100 only two regions are allowed: one around $\mhalf=400$ GeV and
one around $\mhalf=180$ GeV.  Only the latter is compatible with the
EGRET excess. It requires $\mzero$ to be above 1 TeV, which yields
squark and slepton masses above 1 TeV. The gluinos and charginos are
relatively light, as shown on the right hand side of Fig.
\ref{relic}. It should be noted that the EGRET data combined with
the WMAP cross section select basically a single point in parameter
space. Compared with scans over the multidimensional SUSY parameter
space, even with millions of points, it is very easy to miss such a
single point, as demonstrated by recent
scans\cite{ellis,roszkowski}, which missed the EGRET point.

\section{Summary and Outlook}
In summary, the EGRET data shows an intriguing hint of DM
annihilation, since it explains many unrelated facts simultaneously:

a) An excess of diffuse galactic gamma rays which shows a {\it
spectrum} consistent with the expectation from WIMP annihilation
into mono-energetic quarks.

b) The excess is present in {\it all} sky directions with the same
spectrum, thus excluding that it originates from anomalous
contributions in the centre of the galaxy.

c) The excess shows a strongly increased intensity at positions
where extra DM is expected, namely at two doughnut shaped structures
at radii of 14 and 4 kpc from the centre of the galaxy. At 14 kpc
one has observed a ring of stars thought to originate from the
infall of a dwarf galaxy, while at 4 kpc one finds an enhanced
concentration of molecular hydrogen thought to form from atomic
hydrogen in the presence of dust or heavy nuclei, which can be
collected in the gravitational potential of a ring of DM.

d) The enhanced excess of gamma rays cannot be due to additional gas
in these rings as proven by the rotation curve calculated from the
gamma ray excess: the mass in the rings perfectly describe the
hitherto unexplained change of slope in the rotation curve at a
distance of about 11 kpc. The amount of  visible matter is far too
low to have such an impact on the rotation curve.

In this analysis only  the known spectral shapes of the various
processes with arbitrary normalizations are fitted, so  the analysis becomes
largely model independent. Interestingly,  the normalization factors
come out to be in excellent agreement with expectations, both for
the WIMP signal and the background.

Alternative models trying to explain the EGRET excess have to assume
that the locally measured fluxes of protons and electrons are not
representative for our galaxy, in which case these spectra  outside
our local bubble can be tuned to obtain the more energetic gamma
rays needed for the EGRET excess, although these models provide
significantly worse fits to the data, if one takes the strong
correlations in the errors between the different energy bins into
account. In addition such models cannot explain simultaneously the
stability of the ring of stars at 14 kpc and the change of slope in
the rotation curve at $r\approx 11$ kpc.

The results mentioned above make no assumption on the nature of the
Dark Matter, except that its annihilation produces hard gamma rays
consistent with the fragmentation of mono-energetic quarks between 50
and 100 GeV. WIMP masses in this range and the observed
WIMP self annihilation cross section  are consistent with WIMP's being the
Lightest Supersymmetric Particle predicted in the Minimal
Supersymmetric Model with supergravity inspired symmetry breaking,
called the mSUGRA model.

Within this supersymmetric model one finds a spin-independent cross
section for elastic scattering of a WIMP on a proton of about
$10^{-43}~\rm cm^2$, which is within reach\cite{gait} of future
experiments as shown in Fig. \ref{direct}. This elastic scattering
cross section was calculated with Darksusy\cite{darksusy}.

Direct and indirect detection experiments do not prove the
supersymmetric nature of the WIMP's. If the WIMP's are indeed the
lightest supersymmetric particle, then this will become clear at the
future LHC collider under construction at CERN in Geneva, where
supersymmetric particles of the mass range deduced from the EGRET
data should be observable from 2008 onwards, if they
exist.

The statistical significance of the EGRET excess of at least 10
$\sigma$ combined with  all   features mentioned above provides an
intriguing hint that DM is not so dark, but visible by its annihilation.

I thank my close collaborators A. Gladyshev, D. Kazakov, C. Sander
and V. Zhukov for their contributions to this interesting project.
Furthermore  I thank V. Moskalenko, A. Strong and  O. Reimer  for
numerous discussions on galactic gamma rays.

 This work was supported by the DLR
(Deutsches Zentrum f\"ur Luft- und Raumfahrt)
 and a grant from the
DFG (Deutsche Forschungsgemeinschaft, Grant 436 RUS 113/626/0-1).

\end{document}